\documentclass[11 pt]{amsart}
\usepackage[colorlinks=true, pdfstartview=FitV, linkcolor=blue, citecolor=blue, urlcolor=blue]{hyperref} 
\usepackage[table,RGB]{xcolor}
\usepackage{lineno}
\usepackage{graphicx, multicol,changepage,tikz}
\usepackage{tabularx,caption,geometry}
\usepackage{hhline,multirow}
\usepackage{scalefnt}
\makeatletter
\renewcommand{\maketitle}{%
  \bgroup%
  \setlength{\parindent}{0pt}%
  \begin{flushleft}%
    {\Large\textbf{\@title}}%
    \vspace{0.3cm}%
    \newline%
    {\Large {\@author}}%
  \end{flushleft}%
  \egroup%
}
\makeatother
\title{A Reanalysis of the FDA's Benefit--Risk Assessment of Moderna's mRNA-1273 COVID Vaccine Based on a Model Incorporating Benefits Derived from Prior COVID Infection}
\newlength{\extralength}
\definecolor{mypink2}{RGB}{255, 205, 203}
\definecolor{SpringGreen}{RGB}{158, 255, 158}
\definecolor{mygray}{gray}{0.8}
 \definecolor{HighlightYellow}{RGB}{241, 231, 64}
\numberwithin{equation}{subsection}
\newcommand{\orcidicon}{\includegraphics[width=0.32cm]{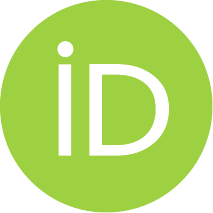}}

\foreach \x in {A, ..., Z}{%
\expandafter\xdef\csname orcid\x\endcsname{\noexpand\href{https://orcid.org/\csname orcidauthor\x\endcsname}{\noexpand\orcidicon}}
}

\begin{document}

\thispagestyle{empty}
\vspace{-.2in}

\begin{center}{\large \bf  \parbox{6.5in}{ 
 A Reanalysis of the FDA's Benefit--Risk Assessment of Moderna's mRNA-1273 COVID Vaccine Based on a Model Incorporating Benefits Derived from Prior COVID Infection
}}

\vspace{.1in}

\parbox{6.2in}{Paul S.\ Bourdon$^1$\orcidA{}, PhD;   Ram Duriseti$^2$\orcidB{}, PhD, MD; H.\ Christian Gromoll$^{1}$\orcidC{}, PhD; Dyana K.\ Dalton$^{3}$\orcidE{}, PhD; Kevin Bardosh$^{4}$, PhD; and Allison E.\  Krug$^{5}$\orcidG{}, MPH}
\smallskip

.\begin{minipage}{6in}{\footnotesize
$^{1}$ \parbox[t]{5.9in}{Department of Mathematics, University of Virginia, Charlottesville, VA,  USA}\\[3pt]
$^2$ \parbox[t]{5.9in}{Stanford Department of Emergency Medicine, Stanford, CA, USA}\\[3pt]
$^{3}$  \parbox[t]{5.9in}{Immunology and Molecular Biology, Independent Research, Hanover, NH, USA} \\[3pt]
$^4$   \parbox[t]{5.9in}{Kellogg's College, University of Oxford, Oxford, UK}\\[3pt]
$^5$ \parbox[t]{5.9in}{Epidemiology, Independent Research,  Virginia Beach, VA, USA}}
\end{minipage}
\end{center}
\smallskip

{\small: Correspondence:  psb7p@virginia.edu}
\bigskip

{\small \textbf{Mathematics Subject Classification 2020}: 92-10, 92C50}

\vspace{.2in}

{

\begin{center}
\begin{minipage}{5.6in}
\noindent{{\bf  Abstract.}\\[4pt] \small
{\bf Background:}  The U.S.\ Food and Drug Administration (FDA) conducted a benefit--risk assessment for Moderna’s COVID vaccine mRNA-1273 prior to its full approval, announced 31 January 2022.  The FDA's assessment focused on males 18--64 years old because its risk analysis was limited to vaccine-attributable myocarditis/pericarditis (VAM/P), given the excess risk among males. The FDA's analysis concluded that vaccine benefits outweighed risks, even for 18--25-year-old males (those at highest VAM/P risk).  We reanalyze the FDA's benefit--risk assessment using information available through the third week of January 2022 and focusing on 18--25-year-old males.\\[1pt]
{\bf Methods:} We develop a benefit--risk model, extending the FDA's, that can stratify benefits and risks of vaccination by prior-infection and comorbidity status. We use the FDA's framework but apply our model to account for benefits derived from prior COVID infection, while also accounting for finer age stratification in COVID-hospitalization rates, incidental hospitalizations (those of patients who test positive for COVID but receive treatment for something else),  more realistic projections of Omicron-infection rates, and more accurate VAM/P rates.\\[1pt]
{\bf Results:} With hospitalizations as the principal endpoint of the analysis (those prevented by vaccination vs.\ those caused by VAM/P), our model finds vaccine risks outweighed benefits for 18--25-year-old males, except in scenarios projecting  implausibly high Omicron-infection prevalence.  Our assessment suggests that mRNA-1273 vaccination of 18--25-year-old males generated between 8\% and 52\% more hospitalizations for VAM/P compared to COVID hospitalizations prevented (over a five-month period of vaccine protection assumed by the FDA). The preceding assessment uses model inputs based on data available at the time of the FDA's mRNA-1273 assessment. Moreover, these inputs as well as model outputs are validated by subsequently available data.\\[1pt] 
{\bf Conclusions:}  The outcome of a vaccine benefit--risk assessment may be dramatically impacted by accounting for the benefits derived  from prior infection by the vaccine-targeted disease.  To increase public confidence in vaccines and thereby reduce vaccine hesitancy,  public-health agencies should employ benefit--risk models capable of supporting stratification of vaccination recommendations not only based on age and sex but also on prior-infection and comorbidity status. }
\end{minipage}
\end{center}

\smallskip

}

\section{Introduction}
Because vaccines can be given to large populations of healthy persons and can be introduced as mandatory,   robust benefit--risk analyses  must be conducted to ensure that vaccines have highly favorable benefit--risk profiles \cite{Arlegui}. 

The Food and Drug Administration (FDA) issued emergency-use authorizations for mRNA COVID-19 vaccines from Pfizer (BNT162b2) and Moderna (mRNA-1273) in December 2020 \cite{Fortner}. Once the mRNA COVID-19 vaccines were given to large populations, it became evident that they could cause myocarditis or pericarditis (hereafter myo/pericarditis), inflammation of the heart or of the membrane surrounding the heart, respectively, with studies indicating the highest risk among males under 40 years old \cite{Buchan, Sharff, Karlstad, Patone, Vu, Naveed, Knudsen}.  Possible mechanisms by which mRNA COVID vaccination causes myocarditis are explored in  \cite{Fanti, Cao}. 

The FDA published two benefit--risk analyses of primary series mRNA vaccination, the first for BNT162b2 \cite{Funk} and the second for mRNA-1273 \cite{Yogurtcu}.    To assess vaccine benefits, the FDA estimated vaccine-preventable COVID-19 cases, hospitalizations, intensive-care unit (ICU) admissions, and deaths \cite{Funk, Yogurtcu}.  To assess vaccine risks, the FDA estimated vaccine-attributable myo/pericarditis (VAM/P) cases, hospitalizations, ICU admissions, and deaths \cite{Funk, Yogurtcu}. The FDA concluded that the benefits of the mRNA vaccines outweighed VAM/P risk even for males in its highest-risk age groups: 16--17 years for BNT162b2~\cite{Funk} and 18--25 years for mRNA-1273 \cite{Yogurtcu}. To simplify our exposition, we  focus mainly on the FDA's analysis for Moderna’s mRNA-1273---there are significant differences in modeling assumptions in the FDA's Pfizer and Moderna assessments. Also, VAM/P risk is substantially greater for mRNA-1273 than for BNT162b2  \cite{Buchan, Sharff, Karlstad, Patone, Vu, Naveed, Knudsen}.

Our benefit--risk modeling differs from the FDA's primarily by its accounting for the protection conferred by prior infection, and, at the time of the FDA’s Moderna analysis (mid-January 2022), likely over 70\% of 18--25-year-olds in the U.S. had been infected by COVID-19---see Supplement S1.   On the risk side, we use data  available at the time of the FDA's mRNA-1273 assessment, including that from \cite{BuchanPreprint, PatonePreprint, SharffPreprint,Funk,Klein}, suggesting VAM/P risk for mRNA-1273 exceeded FDA estimates; see Section~\ref{VAMPRisk} below.

Here, our principal aim is to demonstrate how to incorporate the protection derived from prior infection into a population-level quantitative benefit--risk model for a two-dose vaccination series.  Our model has hospitalizations as its endpoint, comparing the hospitalizations-prevented benefit of vaccination to the hospitalization risk attributable to vaccination.   We illustrate the utility of our model by applying it to extend and improve the FDA's mRNA-1273 assessment relative to its hospitalizations endpoint, which can be viewed as the most important of the agency's endpoints owing to the public-health concern that during waves of the COVID pandemic hospitals might be so overwhelmed that they would not be able to deliver high quality care to their patients. 

 In addition to prior-infection protection, our modeling accounts for incidental hospitalizations, age-specific infection-hospitalization rates, and a more accurate assessment of VAM/P risk.  We apply our model using data available to the FDA at the time of its benefit--risk analysis for mRNA-1273, assumed to be the 3rd week of January 2022 (see Section~\ref{TOA}). We present evidence, using real-world data, that our model's estimates of the hospitalizations-prevented benefit of mRNA-1273 vaccination are significantly more accurate than those the FDA obtained in \cite{Yogurtcu}.   We also analyze benefits and risks of vaccination for specific subpopulations such as males 18--25 having prior-infection protection and no comorbidities.   Our model can be adapted to produce a benefit--risk assessment for any vaccine designed for deployment on a population-wide scale, stratified not only based on age and sex but also on prior-infection and comorbidity status.
 
We emphasize that the application of our model to reanalyze the FDA's benefit--risk assessment for mRNA-1273 is for demonstration purposes and does not constitute a comprehensive re-evaluation, which would address all of the FDA's benefit--risk endpoints (cases, hospitalizations, ICU admissions, deaths).

\subsection{Framework of the FDA's  {\rm m}RNA-1273 Benefit--Risk Assessment}\label{ModernaBR}
In  \cite{Yogurtcu}, the FDA describes its benefit--risk modeling of Moderna's mRNA-1273 for males ages 18 to 64 years.   Outcomes are provided via six scenarios reflecting uncertainties in three major model inputs: COVID-19 incidence, vaccine effectiveness against cases ($VE$) as well as hospitalization ($VEH$),  and vaccine-attributable myo/pericarditis (VAM/P) rate.  At the time of the FDA's mRNA-1273 analysis, the CDC was recommending a 5-month interval between primary-series completion and a booster dose. Consequently, the FDA evaluated benefits and risks over a 5-month period, which we take to be 1 January 2022--31 May 2022. 
  The FDA assumed that all vaccine effectiveness estimates remained constant over this 5-month {\it evaluation period}.  
The FDA's major modeling assumptions are summarized in Table~\ref{MIA} below.

   \begin{table}[h!] {\fontsize{10}{11}\selectfont
      \captionsetup{width=0.96\textwidth}
 \caption{\small The FDA's mRNA-1273 model-input assumptions for 18--25-year-old males \cite[Table 1]{Yogurtcu}. }\label{MIA}
\vspace{-0.1in}

\begin{tabular}{|c|c|c|c|}
\hline
\rowcolor{mygray}
Scenario \rule[-15pt]{0 pt}{.7pt}  & COVID-19 Incidence  & Vaccine Effectiveness  & \parbox[t]{2in}{VAM/P Rate Per Million\\ 2nd doses \cite[Table 2]{Yogurtcu}}\\ \hline
 $1$ & \parbox[t]{1.5in}{ Highest 2021 Incidence,  12/25–12/31} & \parbox[t]{1.4in}{Omicron Dominant; \\VE = 30\%, VEH=72\%} & \parbox[t]{2in}{128 cases, 110 hospitalizations} \\  \hline 
  2 & \parbox[t]{1.5in}{Average 2021 Incidence  \rule[-6pt]{0in}{.25in}}  &  \parbox[t]{1.4in}{Same as Scenario 1} &  \parbox[t]{1.4in}{Same as Scenario 1}\\ \hline
   3 &\parbox[t]{1.5in}{Lowest 2021  Incidence, 6/5  \rule[-5pt]{0in}{.2in}\rule[-5pt]{.1in}{0in} } & \parbox[t]{1.4in}{Same as Scenario 1}  & \parbox[t]{1.4in}{Same as Scenario 1}  \\ \hline
   4 & \parbox[t]{1.4in}{Same as Scenario 1}  & \parbox[t]{1.4in}{Delta Dominant;\\ VE = 80\%, VEH=90\%} & \parbox[t]{1.4in}{Same as Scenario 1}  \\  \hline 
   5 & \parbox[t]{1.4in}{Same as Scenario 1} &  \parbox[t]{1.4in}{Same as Scenario 1} & \parbox[t]{2in}{68 cases, 58 hospitalizations} \\  \hline 
   6 & \parbox[t]{1.4in}{Same as Scenario 1} &  \parbox[t]{1.4in}{Same as Scenario 1} & \parbox[t]{2in}{241 cases, 207 hospitalizations} \\  \hline 
     \end{tabular}}   
     \end{table}

Using inputs from the first row of Table~\ref{MIA} and incidence data from \cite[Table 3]{Yogurtcu},  the FDA concluded for Scenario 1, its ``most likely scenario,'' that vaccinating one million males ages 18--25 years with two doses would prevent, over 5 months, 82,484 COVID-19 cases, 4766 COVID-19 hospitalizations, 1144 ICU admissions, and 51 deaths due to COVID-19, while risking 128 VAM/P cases, 110 VAM/P hospitalizations, 0~ICU admissions and 0~deaths from VAM/P \cite[Table 2]{Yogurtcu}.  Observe that the FDA does include a high VAM/P-risk scenario (6) as well as lower-benefit scenarios (2 and 3).

\subsection{The Time of the FDA's mRNA-1273 Analysis} \label{TOA}  Although the FDA does not define in \cite{Yogurtcu} the time of its benefit--risk analysis for mRNA-1273, a natural choice is the 3rd week of January 2022, which concluded 10 days before the FDA's announcement of the full approval of mRNA-1273 on 31 January 2022 \cite{FDAModernaPR}.   This choice is consistent with the FDA’s time-of-analysis definition for its BNT162b2 assessment: the 2nd week of August 2021 \cite[Section 2.1]{Funk}, which concluded nine days before the  FDA’s announcement  of the  full approval of BNT162b2 on
23 August 2021 \cite{FDAPressRPfizer}.   Consequently, our reanalysis of the FDA's benefit--risk assessment \cite{Yogurtcu} is based on model inputs derived from information available before 22 January 2022.     Subsequent information validating our input choices will often be provided but only via footnotes. Starting in Section~\ref{SPABMI}, we discontinue basing our modeling on information available only  before 22 January 2022; in Section~\ref{SPABMI},  our intent is to show how our model can be modified to incorporate a comorbidity (obesity) as a risk modifier.

\subsection{The FDA's mRNA-1273 Benefit Assessment} \label{Benefits}

The FDA's computation in \cite{Yogurtcu} of the cases-prevented and hospitalizations-prevented benefits of vaccination is  straightforward.  For each scenario, CDC data is used to anticipate over the 5-month evaluation period a certain number of COVID cases $C$ and hospitalizations $H$ per million unvaccinated males, and the corresponding evaluation-period benefits per 1 million vaccinations are estimated to be 
\begin{equation*}
C\cdot VE\ \text{cases prevented and} \  H\cdot VEH\ \text{hospitalizations prevented},
\end{equation*}
where $VE$ and $VEH$ are, respectively, the effectiveness of full mRNA-1273 vaccination against cases and against hospitalization; the FDA assumed that for Omicron $VE = 0.30$ and $VEH= 0.72$, as noted in Table \ref{MIA} above.

 For Scenario 1,  its most likely scenario, and two others, the FDA assumed that peak 2021 Omicron-case rates  (late December 2021) would be sustained for 5 months, resulting in an exceptionally high projection of COVID-19 cases among the unvaccinated; e.g., $C$ = 274,947 for 18--25-year-old males. Similarly, for these scenarios, the FDA assumed that the CDC's reported late-December 2021 Omicron-hospitalization rate for unvaccinated \mbox{18--45-year-old} males would be sustained for 5 months and assumed that the same rate would apply to the younger 18--25 bracket, arriving at \mbox{$H=6619$ hospitalizations}. Thus, for these scenarios, the FDA predicted that full mRNA-1273 vaccination of 1 million \mbox{18--25-year-old} males would prevent 274,947$\cdot 0.30 \approx$ 82,484 COVID cases and \mbox{$6619 \cdot 0.72 \approx 4766$} COVID hospitalizations over the 5-month evaluation period  \cite[Table 2]{Yogurtcu}.

 {\em How our benefit assessment differs from the FDA's:}

$\bullet$ {\em We account for the protection conferred by prior infection.} At the time of the FDA's mRNA-1273 assessment, there was substantial evidence that prior infection would provide significant protection against COVID infection and hospitalization. In a study published \mbox{5 January 2021}, Dan et al.\ found ``About 95\% of subjects retained immune memory at $\sim$6 months after [COVID] infection'' \cite{Dan}.  A meta-analysis  of studies published December 2020 through August 2021 demonstrated that protection from prior COVID infection ``is, at least, equivalent to the protection afforded by complete vaccination of COVID-na\"ive populations'' \cite{Shenai}. The CDC acknowledged in a 29 October 2021 science brief  that ``The immunity provided by vaccine and prior infection are both high'' \cite{CDCSB},  while another  CDC study found that during the Delta wave, unvaccinated persons with prior COVID infection had significantly greater protection against infection and hospitalization than did fully vaccinated, COVID-na\"{i}ve persons \cite{Leon}.  Moreover, in December 2021, the FDA acknowledged, ``The omicron variant has significantly more mutations than previous SARS-CoV-2 variants, particularly in its S-gene, the gene that encodes the virus's spike protein''\cite{FDAOM}, suggesting that prior infection, entailing exposure to all SARS-CoV-2 proteins (e.g., the nucleoprotein),  would likely yield greater protection against Omicron than would mRNA-1273 vaccination, which entails exposure only to the  spike protein of the ancestral strain of SARS-CoV-2. Among the findings of a study of SARS-CoV-2 specific B and T cell memory is ``CD8+ T cell responses preferentially target the nucleoprotein, highlighting the potential importance of including the nucleoprotein in future vaccines''  \cite{Cohen}.\footnote{Consistent with the preceding paragraph's evidence, a Qatari study found prior infection provided 52.2\% effective protection against symptomatic Omicron infection and 91\% effective protection against ``severe, critical, or fatal'' Omicron, with corresponding effectiveness estimates for two doses of mRNA-1273 being 2.2\% and 66.3\%, respectively \cite[Figure 2C, 2D]{Qatar}   (study period 23 December 2021--21 February 2022). Findings from the meta-analyses and systematic reviews \cite{BWM, IHMEFT, Song, Feikin} provide convincing evidence that prior infection provided greater protection against Omicron during the evaluation period than did two-dose vaccination among the COVID-na\"{i}ve [Supplement S1, Section S7].}

Further evidence of the need to account for prior-infection protection includes (i) a CDC estimate that 54.9\% of  18--49 year-olds had been infected by COVID-19 by 30 September 2021~\cite{CDC1}, (ii) a South African study \cite{DH} released 14 December 2021 suggesting overall effectiveness of prior- infection protection against Omicron to be about 47\% [Supplement S1, Section S4],  and (iii) data from England \cite{Ferguson} released 22 December 2021 suggesting that prior infection among the unvaccinated provides significant protection from Omicron hospitalization [Supplement S1, Section S5].

$\bullet$ {\it We account for incidental COVID-19 hospitalizations.}  At the time of its mRNA-1273 assessment, incidental-hospitalization rates were well known to be substantial.  On  9 January 2022, the director of the CDC stated, ``[U]p to 40 percent of the patients who are coming in with COVID-19 are coming in $\ldots$ with something else'' \cite{Walensky}.\footnote{We searched unsuccessfully for published CDC incidental-COVID-hospitalizations data. A large U.S.\ study suggests  an incidental rate for 0--18 year-olds typically well above 50\% during its study period 1 March 2020--31 January 2022 and for those 19--64 an incidental rate above 40\% for 1 January 2022--31 January 2022~\cite[Figure2]{Epic}.}   On 10 January 2022, Ontario began publishing incidental rates, with rates for January 8--15 exceeding 45\% \cite{PHO_IH}. There is evidence that incidental rates are especially high for the young. For example,  a major study \cite{Sorg} found,  ``As of May 2021, the cumulative rate for hospitalization associated with SARS-CoV-2 infection was 35.9 per 10,000 children'' and ``When limiting the analysis to include only patients with COVID-19 who required therapeutic interventions, the hospitalization rate decreased 5.5-fold to 6.5 per 10,000 children,'' which suggests an incidental rate of over 80\% ``for children'' (persons 0--17-years-old). See Supplement S1 for additional incidental-hospitalization~data.

$\bullet$  {\it We account for finer age stratification in COVID-hospitalization rates.}   The FDA assumed that hospital admission rates were uniform for males of ages 18--45, contradicting CDC models indicating that hospitalization rates for those 30–49 are \mbox{two times} the rates for 18–29-year-olds \cite{CDC2}.  In contrast, we employ an Omicron infection-hospitalization rate estimate for precisely the group of interest---males, 18--25 years old.  Our estimate, based on a model developed in \cite{HEP}, is consistent with COVID-hospitalization data for the U.S. military [Supplement S1, Section S2].
 
$\bullet$ {\it  We use a CDC case-to-infections multiplier to inform our projections of Omicron-infection prevalence.} The FDA assumed in its most likely scenario (as well as Scenarios 5 and 6) that case and hospitalization rates would remain constant at peak 2021 Omicron levels for 5 months.  Throughout the pandemic, the number of COVID infections has always been significantly larger than the corresponding number of reported cases.  For the period February 2020 through September 2021, the CDC estimated the number of infections to be four times the corresponding number of reported cases \cite{CDC1}. Moreover, the case-to-infection multiplier likely increased  (e.g., owing to increased home testing\footnote{See \cite{IHT}})  during the Omicron-dominant period in the U.S., which started  mid-December 2021.    If we conservatively  assume that the CDC's 4:1 infection:case ratio holds over the evaluation period, we find the FDA's evaluation-period case-count projection of 274,947 per 1 million unvaccinated 18--25-year-old males (Scenarios 1, 5, 6)  corresponds to 4 · 274,947 = 1,099,788 infections, well over 1 million. Rounding down, we deem the FDA's case-total assumption to be equivalent to the assumption that among one million unvaccinated males, there would be, over 5 months, 1,000,000 infections. Having well over one million infections in a group of 1 million 18--25-year-olds over 5 months is implausible. Multiple infections over a 5-month period would be unlikely\ given the high protection from reinfection resulting from recovery from a recent  infection, and any reinfections would be offset by individuals who did not become infected at all (e.g., owing to prior-infection protection, limited exposure to infectious persons, or, possibly, genetic factors \cite{HorowitzPreprint}).\footnote{The preprint \cite{HorowitzPreprint}, posted 10 June 2021, was later published \cite{Horowitz}. Even direct inoculation with  SARS-CoV-2 does not necessarily produce infection: in the challenge study \cite{ICLCT}  of the 36 inoculated, 16 ``remained uninfected.''  }  {\em Thus, for 3 of the FDA's 5 Omicron-based scenarios, including its most likely scenario, the FDA essentially assumed that every unvaccinated male 18--25 would be infected by COVID over the  evaluation period.  }

 In contrast, for our most likely scenario, we project COVID incidence over the evaluation period to be twice that of the 2nd COVID wave in the U.S.  (1 October 2020--28 February 2021), arriving at an estimate that 45.6\% of unvaccinated, COVID-na\"{i}ve 18–25 year-old males in the U.S.\ would be infected by Omicron over the evaluation period---see Section S1 of Supplement S1.\footnote{Based on seroprevalence and modeling information, we estimate the actual evaluation-period infected rate was less than 50\% [Supplement S2, Section S2.3]. }

\subsection{The FDA's mRNA-1273 Risk Assessment} \label{VAMPRisk}

The FDA relied on its Biologics Effectiveness and Safety (BEST) System  for the VAM/P-case rates appearing in Table~\ref{MIA} above and used the CDC's Vaccine Safety Datalink (VSD) system to estimate the percentage of cases requiring hospitalization \cite[Section 2.3.2]{Yogurtcu}. 

VAM/P-case rates vary considerably depending on the case definition employed and the extent of stratification of risk, based on age, sex, vaccine manufacturer, and dose, with most cases occurring postdose 2 \cite{Knudsen}.    For its mRNA-1273 assessment, the FDA's four BEST-System data partners reported, with wide confidence intervals, the following myo/pericarditis incidence rates for 18--25-year-old males during the 7-day interval following dose 2: approximately 35, 70, 146, and 256 cases per million 2nd \mbox{doses \cite[p,\ 12]{Wong}}. The FDA's BEST System uses health-claims data  not validated by a complete review of patients’ medical charts \cite[Section~4]{Yogurtcu}  and has ``inherent limitations, such as small sample sizes and imperfect sensitivity of ICD-10 codes to identify these rare outcomes'' \cite[Section~4]{Funk}.   We rely on four sources of data to estimate the VAM/P rate for dose 2 among 18--25-year-old males: (i) Buchan et al.\ \cite{BuchanPreprint}, who apply the Brighton Collaboration's case definition, (ii) Patone et al.\ \cite{PatonePreprint}, who consider only hospitalized cases of myocarditis, (iii)~Sharff et al.\ \cite{SharffPreprint}, who apply the CDC's case definition \cite{Gargano}, and (iv) the FDA (with VSD-based~adjustments).\pagebreak

  {\it How our mRNA-1273 VAM/P-risk assessment differs from the FDA's:}

$\bullet$ {\it We include dose-one VAM/P events.}  
For its mRNA-1273 analysis \cite{Yogurtcu}, the FDA computed the evaluation-period benefits of {\em two doses} of mRNA-1273 without accounting for the VAM/P risk associated with dose 1, potentially biasing the FDA's analysis in favor of vaccination.  In contrast, for BNT162b2-related VAM/P, the FDA added excess myo/pericarditis cases postdose 1 and postdose 2 to approximate the population-level risk of completing 1 million two-dose vaccination series \cite[Section 2.3.1]{Funk}.  Hence, for BNT162b2, the FDA did account  for 1st-dose VAM/P risk  but did not assess the benefits of dose 1 alone during the interval starting with dose 1 and ending at the point of full vaccination, potentially biasing the analysis against vaccination. Because we, like the FDA, do not account for adverse events of vaccination other than VAM/P, and mRNA vaccination might have unknown long-term adverse health impacts, we believe that even in a benefit--risk analysis assessing benefits only postdose 2, the choice to include 1st-dose VAM/P risk is appropriate, being consistent with the first-do-no-harm principle, the FDA's choice in its BNT162b2 assessment, and the FDA's choice in the benefit--risk analysis of its BLA (Biologics License Application) Clinical Review Memorandum for mRNA-1273---see Section~\ref{BLAScenario} below.   Thus, we include the 1st-dose VAM/P risk for mRNA-1273 in our main analysis.  See Section S2 of Supplement S2 for a sensitivity analysis showing that accounting for dose 1's benefits does not substantially change our benefit--risk conclusions.  

$\bullet$ {{\it  We incorporate evidence that VAM/P risk for mRNA-1273 was over twice that for BNT162b2.}   In a ``Review Memorandum Addendum'' \cite{RMA}, dated 19 November 2021, the FDA describes  evidence that VAM/P risk for mRNA-1273 is significantly greater than that for BNT162b2, including regulatory-agency reports from five countries, data from Ontario later analyzed by Buchan et al.\ \cite{BuchanPreprint}, and data from the CDC: ``[T]he most recent data presented by the Center for Disease Control and Prevention (CDC) analyzed from the Vaccine Safety Datalink (VSD) indicates a higher rate of myocarditis with the Moderna COVID-19 Vaccine than the Pfizer-BioNTech COVID-19 \mbox{Vaccine \ldots}.''}

The ``most recent data'' from the VSD to which the FDA refers was presented at the CDC's October 2021 ACIP meeting \cite{Klein}. Adjusted-rate ratios for 18--39-year-old males from the table on page 27 of \cite{Klein}  suggest the 2nd-dose VAM/P rate for mRNA-1273 is 2.26 times that of BNT162b2. Data from other sources suggests higher rate ratios, e.g., a 2nd-dose rate ratio of 5.1 was reported by  Buchan et al.\   for males 18–24 \cite[p. 5]{BuchanPreprint} and of 5.2 was reported by Husby et al.\ for males 12--39 \cite[paragraph after Figure 3]{Husby}.  A study by Patone et al.\ of myocarditis cases requiring hospitalization within a 28-day risk interval after a vaccine dose found {\it among males under 40} a rate of 101 excess myocarditis hospitalizations per million 2nd doses of mRNA-1273 and 12  per million 1st doses, with corresponding BNT162b2 rates being 12 per million 2nd doses and 3 per million 1st doses  \cite[p.\ 4]{PatonePreprint}.

  Based on ``preliminary findings''  \cite[p.\ 2]{RMA}, having ``large uncertainty'' owing to ``small numbers of observed events''  \cite[p.\ 16]{Wong}, as well as omitting dose-1 cases, the FDA  suggests  that  for 18--25-year-old males, VAM/P risk for mRNA-1273 is comparable to that for BNT162b2:  128 cases per million 2nd doses of mRNA-1273  \cite[Table 4]{Yogurtcu}  vs.\ 131 cases per million full vaccinations of BNT162b2 (among 18--24-year-old males)  \cite[Table 3]{Funk}, with corresponding hospitalization projections of 110 per million 2nd doses of mRNA-1273 and 131 per million full BNT162b2 vaccinations.  {\it Throughout this paper, we assume that VAM/P rates for males 18--24 well approximate those for males 18--25}.

Using the rate ratios and BNT162b2 VAM/P-risk assessment discussed above, we compute a VAM/P-hospitalization rate of 224 per million 2nd doses of mRNA-1273 for males 18--25 years old  [Supplement S1, Section S6.1].
 
    The FDA's BEST System also provided evidence that VAM/P risk for mRNA-1273 might be over twice that for BNT162b2.   An FDA briefing document for a meeting held 14 October 2021 indicates that one of the agency's BEST data partners reported a postvaccination myo/pericarditis incidence rate of 283.7 (95\% CI 145.2--573.5) cases per million full mRNA-1273 vaccinations (7 day risk window after each dose) \cite[Section 2.3.3]{EUABD}.  Applying a background rate of four cases per million per week for 18--25-year-old males for each dose  [Supplement S1, Section S6.3], we find the VAM/P-case rate corresponding to 283.7 is 275.7 per million full vaccinations. Assuming a case-hospitalization rate of 100\%, as in FDA's BNT162b assessment, we obtain 275.7 VAM/P hospitalizations per million full mRNA-1273 vaccinations vs.\ 131 per million full BNT162b vaccinations.

As we have indicated, we rely on four sources of data for our VAM/P-rate estimates for dose 2 of mRNA-1273:  the FDA with VSD-based adjustments, as well as  studies by Buchan et al.\  \cite{BuchanPreprint}, Patone et al.\   \cite{PatonePreprint}, and Sharff et al.\ \cite{SharffPreprint} (from, respectively, Ontario, England, and the U.S.).  The conclusion of \cite{SharffPreprint}, which includes the following, is noteworthy:\footnote{The preprint \cite{SharffPreprint} and the later publication \cite{Sharff} share the same conclusion.}

\begin{quotation}
{\noindent We identified additional valid cases of myopericarditis following an mRNA vaccination that would be missed by the VSD's search algorithm $\ldots$.  The true incidence of myopericarditis is markedly higher than the incidence reported to US advisory committees. }
\end{quotation}

 In Table~\ref{VAMPTable} below, we compare the FDA's projected  mRNA-1273 VAM/P case and hospitalization rates  to those of the other sources discussed above. 
We exclude VAM/P-rate sources, such as \cite{Oster}, depending on strictly passive surveillance---see Section S6.8 of Supplement S1.

\begin{table}[h!] {\fontsize{10}{11}\selectfont
 \begin{adjustwidth}{-1.55cm}{}
\captionsetup{width = 1.1\textwidth}
\caption{\small Estimated VAM/P Events Per Million 2nd Doses of mRNA-1273 Among Males 18--24 or 18--25 years old.$^a$ Based on Data Available Before 1 January 2022}\label{VAMPTable}
\vspace{-.1in}

\begin{tabularx}{1.15\textwidth}{|c|c|X|X|X|X|X|}
\hline
\rowcolor{mygray}
 \hspace{-.1in} Description  & \begin{tabular}{c} FDA\\ Scenarios 1--4\\(Scenarios 5 \&  6) \\ \cite[Table 2]{Yogurtcu}\end{tabular} &       \begin{tabular}{c} \hspace{-1em} Buchan et al.$^b$\\\cite{BuchanPreprint}  \end{tabular} &   \begin{tabular}{c}Patone et al.\\ \cite{PatonePreprint,BuchanPreprint}\end{tabular}&   \begin{tabular}{c} Sharff et al.\\\cite{SharffPreprint}\end{tabular}&\begin{tabular}{c}\hspace{-1em} Data From\\ \hspace{-0.2em} FDA \& VSD \\ \hspace{-0.2em} \cite{Funk, Wong, Klein}\end{tabular} &  \begin{tabular}{c}  \hspace{-0.2em} Weighted \\ \hspace{-0.2em} Average of\\ \hspace{-0.2em}  Highlighted\\  \hspace{-0.2em} Data\end{tabular}\\ \hline
\begin{tabular}{c}Cases\\per millon\end{tabular} &  \begin{tabular}{c}128\\(68 \& 241)\end{tabular} & \cellcolor{HighlightYellow} \rule{.23in}{0in} 302$^c$
& \cellcolor{HighlightYellow} \rule{.23in}{0in} 269 &\cellcolor{HighlightYellow} \rule{.23in}{0in} 525 &\cellcolor{HighlightYellow}  \rule{.23in}{0in}  260 &\hspace{1.4 em} 301$^\dagger$ \\  \hline
\begin{tabular}{c}  Hospitalizations$^d$\\per million\end{tabular} & \begin{tabular}{c} 110\\ (58 \& 207)\end{tabular}
&\cellcolor{HighlightYellow} \rule{.23in}{0in}  242& \cellcolor{HighlightYellow} \rule{.23in}{0in} 231&\cellcolor{HighlightYellow} \rule{.23in}{0in} 452 &\cellcolor{HighlightYellow}  \rule{.23in}{0in}  224 & \hspace{1.4em}  250$^\ddagger$ \\  \hline
\begin{tabular}{c} Number of events\\ \rule{0.08in}{0in} on which estimate\\   \rule{0.06in}{0.0in} \rule[-10pt]{0in}{0.2in} is based\end{tabular} & \begin{tabular}{c}  21\\ \cite[p.\ 14]{Wong}\end{tabular} &  \begin{tabular}{c} \hspace{-0.8em} 55\\ \hspace{-0.8em} \cite[Table 4]{BuchanPreprint}\end{tabular}
& \rule{.232in}{0in} 23  &\begin{tabular}{c}\hspace{-0.8em} 7\\\hspace{-0.8em} \cite[Table 1]{SharffPreprint}\end{tabular}&  \rule{.233in}{0in}  22 & \begin{tabular}{c}  \hspace{-0.75em}  Total Events\\  \hspace{-0.75em} Columns 3--6\\   \hspace{-0.75em}  107\end{tabular}  \\  \hline
 \end{tabularx} 
 \noindent{\parbox{1.15\textwidth}{\footnotesize{$^a$ For discussions of the data in columns 3--6, see Section S6 of Supplement S1; {\it  in particular, data in column 4 are estimates based on findings from \cite{PatonePreprint} for males under 40, adjusted using an age-based myocarditis-incidence ratio as well as pericarditis data from \cite{BuchanPreprint}}.  To obtain the column-5 VAM/P case rate of 525 per million we applied the background myo/pericarditis rate suggested by \cite{BuchanPreprint}.\\
 $^b$  Buchan et al.\ applied the Brighton Collaboration's case definition for myo/pericarditisx, which reduced the reported  417 post-mRNA-vaccination study-period cases of myo/pericarditis to 297 \cite[p.\ 5]{BuchanPreprint}.\\
 $^c$ This estimated VAM/P-case rate is irrespective of the corresponding 1st-dose type. For homologous dosing (1st dose also mRNA-1273), Table 3 of \cite{BuchanPreprint}  indicates the rate rises to at least 320 per million when the 2nd dose is taken less than 8 weeks after the 1st (with 4--6 weeks being the recommended interval \cite{IR2}).\\
 $^d$ For its BNT162b2 assessment, the FDA assumed all VAM/P cases were hospitalized \cite[Section 2.3.2.2]{Funk}.  For its mRNA-1273 assessment the FDA assumed VAM/P case-hospitalization rates decrease with age, starting with 86\% for males 18--25 and ending with 77\% for males 36 and over \cite[Section 2.3.2.2]{Yogurtcu}.  We apply the 86\% rate in columns 4--6 (which, for column 5, matches the case hospitalization rate: 6 of the 7 cases were hospitalized). For column 3, Buchan et al.\ report a case-hospitalization rate  postdose 2 of an mRNA vaccine (all ages, males \& females) of 79.2\% ($\approx 164/207 \times 100$\%) \cite[Table 1]{BuchanPreprint}; we expect the rate to be higher for the age range 18--24, but, conservatively, choose a hospitalization rate of 80\%.\\
 $^\dagger$   $301 \approx 302\cdot \frac{55}{107} .+ 269\cdot \frac{23}{107} + 525\cdot \frac{7}{107} + 260\cdot \frac{22}{107}$ \hspace{1in} $\ddagger$ $250 \approx 242\cdot \frac{55}{107} .+ 231\cdot \frac{23}{107} + 452\cdot \frac{7}{107} + 224\cdot \frac{22}{107}$ }}}
\end{adjustwidth}}
\end{table}
\subsection{A Seventh Benefit Risk Scenario for mRNA-1273 }\label{BLAScenario}

The FDA's BLA Clinical Review Memorandum \cite{FDABLACR} for mRNA-1273, completed 28 January 2022, describes a benefit--risk scenario  not matching any scenario of the FDA's  published benefit--risk assessment of mRNA-1273   \cite{Yogurtcu}.   The projections of the BLA-Memo Scenario \cite[Section 4.7]{FDABLACR} are described in the initial row of data of Table~\ref{SevenScenarios} below and compared to the FDA's scenarios of \cite[Table 2]{Yogurtcu} in particular to Scenario 1, the most likely scenario.   Differences include the BLA-Memo Scenario's accounting for 1st-dose VAM/P risk and projecting 1 million full vaccinations would cause 47 ICU admissions (vs.\ 0 admissions).\footnote{Subsequent analyses provide further evidence that the scenarios of  \cite{Yogurtcu} may underestimate VAM/P severity; e.g., a  meta-analysis \cite{Yasuhara} of 23 studies including 854 patients ages 12--20 years found a 92.6\% case-hospitalization rate for postvaccination myo/pericarditis and a 23.2\% ICU-admission rate. }

\begin{table}[h] {\fontsize{10}{11}\selectfont
\begin{adjustwidth}{-0.3cm}{}
\captionsetup{width=1.06\textwidth}
\caption{\small FDA scenarios projecting benefit-risk outcomes per million 18--25-year-old males vaccinated with two primary series doses of mRNA-1273.*}  \label{SevenScenarios}
\vspace{-.1in}

\begin{tabular}{|c|c|c|c|c|c|c|c|c|}
\hline
 \cellcolor{mygray} &  \multicolumn{4}{|c|}{  \cellcolor{mygray}  \rule[-6pt]{0in}{0.25in} Benefits (Outcomes Prevented) } &\multicolumn{4}{|c|}{  \cellcolor{mygray}VAM/P Risks}\\ \hhline{|*1{>{\arrayrulecolor[gray]{.8}}-}>{\arrayrulecolor{black}}|*8{-}|}\multirow{-2}{*}{ \cellcolor{mygray} Scenario}   &  \cellcolor{mygray} \rule[-5pt]{0in}{0.2in}\ Cases &  \cellcolor{mygray} Hospitalizations &   \cellcolor{mygray}\begin{tabular}{c}ICU\\ Visits \end{tabular} &  \cellcolor{mygray}  Deaths &  \cellcolor{mygray} Cases  &  \cellcolor{mygray} Hospitalizations&  \cellcolor{mygray} \begin{tabular}{c} ICU\\ Visits\end{tabular} &  \cellcolor{mygray} Deaths\\ \hline
 \rowcolor{HighlightYellow}
BLA-Memo & 76,362
& 1,755& 421 & 4 & 148 & 128 & 47 & 0\\  \hline
\rowcolor{HighlightYellow}
1 & 82,484 & 4,766 & 1,144  &51&  128 & 110 & 0&0 \\  \hline
2 & 26,705 & 2,088 & 501  &48 &  128 & 110 & 0&0 \\  \hline
3 & 3,903 & 635 & 152  & 7 &  128 & 110 & 0&0 \\  \hline
4 & 219,958 & 5,957 & 1,430  &63 &  128 & 110 & 0&0 \\  \hline
5 & 82,484 & 4,766 & 1,144  &51&  68 & 58 & 0&0 \\  \hline
6 & 82,484 & 4,766 & 1,144  &51&  247 & 207 & 0&0 \\  \hline
 \end{tabular} 
  \noindent{\footnotesize{\ * For Scenarios 1--6, the FDA estimated VAM/P risk only for dose 2.   Projections for the FDA'a Scenarios 1--6 are from \cite[Table 2]{Yogurtcu}. BLA-Memo-scenario projections are from \cite[Section 4.7]{FDABLACR}.   }}
 \end{adjustwidth}}
\end{table}

Consistent with the BLA-Memo Scenario, the FDA's 31 January 2022 press release \cite{FDAModernaPR} announcing the full approval of mRNA-1273  indicates some mRNA-1273 VAM/P cases require ICU care.

To support its estimates of VAM/P-hospitalization and ICU-admission risk, the FDA in \cite{Yogurtcu} cites a source of VSD data \cite{NK2} with a collection-cutoff point of 21 August 2021, over four months before  completion of its mRNA-1273 assessment.  The VSD data from \cite{NK2} cited by the FDA \cite{Yogurtcu}  is  scant, reporting only 21 postvaccination myo/pericarditis cases for those 18--29 \cite[p.\ 16]{NK2}.   A comparison of Buchan et al.'s myo/pericarditis data from Ontario~\cite[Table 4, collected 1 June 2021--4 September 2021]{BuchanPreprint}  to VSD data through 19 October 2021 \cite[pp.\ 8, 16] {Klein} shows that, among males and females ages 18--39, the number of myo/pericarditis cases per million 2nd doses of mRNA-1273 detected by Public Health Ontario is 3.76 times  the number detected by the CDC's VSD system [Supplement S2, Appendix S3].

\section{Methods}  \label{MethodsSec}

Recall that for its cases and hospitalizations endpoints, the FDA's benefit--risk model has inputs $C$ and $H$ (the numbers of anticipated evaluation-period COVID cases and hospitalizations, respectively,  in a test population of one million unvaccinated males) as well as $VE$ and $VEH$ (the fractions of these events prevented by full vaccination). These model inputs are combined to project the cases-prevented and hospitalizations-prevented benefits of vaccinating the test population, and these benefits are compared to  the model's corresponding risk inputs, the numbers of VAM/P cases and hospitalizations projected for the 2nd mRNA-1273 dose.

Our model, which has hospitalizations as its endpoint, is algebraic (like the FDA's) and thus is widely accessible. Our model extends the FDA's by not only incorporating prior-infection protection but also by being essentially equivalent to the FDA's model if prior infection is assumed to provide no protection. Because the FDA relied on CDC data to supply projected values for $C$ and $H$, the FDA's model does not explicitly link cases to hospitalizations and similarly does not explicitly link $VE$ and $VEH$.  In contrast, our model does provide these links.  For example, it links infections, a key model input replacing reported cases, to hospitalizations via another key input---the {\it infection-hospitalization rate} IHR.  Additional input variables are described below.

Model inputs are combined to project the hospitalizations-prevented benefit of vaccinating a test population, and this projected benefit is compared to the model's risk input, the projected number of VAM/P hospitalizations expected to occur in the course of achieving a fully vaccinated test population.   {\it  In our modeling, ''infection'' is any COVID infection, symptomatic or asymptomatic.}

{\em Test population.}
Our model is based on a hypothetical test population of 1 million males aged 18--25.  We assume that the test population is representative, as of 1 January 2022 of all males in this age group in terms of: distribution of ages within the 18--25 range, comorbidities  (e.g., obesity, cardiovascular disease, diabetes), and the percentage having had a prior COVID-19 infection.   We project the number of hospitalizations for Omicron infection that will occur in this test population over the evaluation period 1 January  2022--31 May 2022 (a) assuming all members remain unvaccinated throughout the period, and (b) assuming all members of the population are vaccinated throughout the period (all fully vaccinated, 14 days past dose 2, on 1 January 2022). The difference---the projection of (a) minus that of (b)---is the hospitalizations-prevented benefit of vaccinating the test population.  We call a test population consisting of infection-na\"ive unvaccinated persons a {\it baseline test population}.

\subsection{Model Inputs}\label{ModIn}

{\em Infection rate $I_r$.}
 We base our modeling  on $I_r$, the fraction of a  baseline test population of unvaccinated infection-na\"ive males 18--25 projected to become infected (symptomatic and asymptomatic) over the FDA's 5-month evaluation period. 

{\em Infection-hospitalization rate  $H_r$.}
We translate infections to projected hospitalizations using an estimated infection-hospitalization rate $H_r$ for unvaccinated, infection-na\"{i}ve males 18--25 years old. $H_r$ is the fraction of a population of unvaccinated  individuals experiencing a primary COVID infection  that will progress to severe disease requiring hospitalization.

{\em Prior-infection fraction $F_{pi}$}. This is the fraction of the test population having had a COVID-19 infection before the start of the  evaluation period.

{\em Vaccine effectiveness against infection $E_v$.}  This is the fraction by which the baseline infection rate $I_r$ is reduced due to the protection afforded by vaccination; i.e., the infection rate for a fully vaccinated test population is $I_r(1-E_v)$.

{\em Prior-infection and hybrid effectiveness against infection, $E_{pi}$ and $E_h$.} These are the fractions by which the baseline infection rate $I_r$ is reduced due, respectively, to the protection afforded by prior infection, or to both vaccination and prior infection (hybrid protection). 

{\em Hospitalization-risk reduction $HRR_v$, $HRR_{pi}$, and $HRR_h$}. These are the fractions by which the hospitalization rate $H_r$ is reduced for populations that are protected, respectively, by vaccination, prior infection, or both (hybrid protection).

{\em VAM/P hospitalizations $H_{\rm VAM/P}$.} Our final model input $H_{\rm VAM/P}$ is the projected number of VAM/P hospitalizations expected to occur in attaining a test population of 1 million fully mRNA-1273 vaccinated 18--25-year-old males.  To fully vaccinate 1 million individuals with an mRNA COVID vaccine, more than 2 million doses must be administered (assuming, e.g., that those who experience myo/pericarditis related to dose 1 do not take dose 2), and the adverse events associated with each of the over 2 million doses reflect the cost to the population of obtaining the benefits of 1 million full vaccinations.

\subsection{Combining Model Inputs to Produce Outputs} \label{MIO}
Recall that we are considering a test population of 1 million males 18--25  over the evaluation period 1 January 2022--31 May 2022. Let $U$ denote the expected number of  Omicron hospitalizations occurring in this test population over the 5-month evaluation period assuming its members remain unvaccinated throughout the period.  Let $V$ denote the corresponding number of expected hospitalizations assuming the population is vaccinated throughout the period.  The difference $U -V$ is the hospitalizations-prevented benefit of vaccination for the test population, and the outputs $U$ and $V$ can be written in terms of the model inputs defined in the preceding section  (thus, like the FDA's, our model is deterministic). 

In our modeling, we consider only the possibility of hospitalizations associated with test-population members' first infections during the evaluation period. We expect second infections over a 5-month period to be rare; moreover, infection-hospitalization risk will drop significantly after infection  \cite{Shenai, CDCSB, Leon}.  Our decision to analyze hospitalization risk associated with only first infections during the 5-month evaluation period allows us to assume,  as did the FDA, that all effectiveness factors contributing to the analysis remain constant over the evaluation period.  Each effectiveness value represents average effectiveness over the evaluation period; in this way, our modeling accounts for waning of vaccine and prior-infection protection.

{\em Deriving $U$.} To derive $U$, we will assume that the test population remains unvaccinated during the evaluation period. As a first step, suppose that the entire test population is infection na\"ive. Then, in this baseline population,  we expect $(1{\rm m}) \cdot I_r$ infections to develop over 5 months, resulting in $(1{\rm m}) \cdot I_r\cdot H_r$ hospitalizations, where  $1{\rm m} =$ 1 million.

Next, suppose a fraction $F_{pi}$ of the unvaccinated test population has had a COVID-19 infection  prior to the start of the evaluation period.  We must account for the protection against infection  and hospitalization afforded by prior infection. We accomplish this in 
two stages, first accounting for protection against infection, then for protection from progression to severe disease for infected individuals. The first stage depends on the model input $E_{pi}$, while the second stage uses $HRR_{pi}$. 

So, with fraction $F_{pi}$ of the test population previously infected, $(1-F_{pi}) \cdot (1{\rm m})$  are infection-na\"{i}ve and will respond like the baseline population described above. That is, we expect $(1 - F_{pi}) \cdot (1{\rm m}) \cdot I_r$  infections leading to $(1-F_{pi}) \cdot (1{\rm m}) \cdot I_r \cdot H_r$ hospitalizations from this group. Among the remaining $F_{pi}\cdot (1{\rm m})$ members of the test population with prior infection, we expect $F_{pi}\cdot (1{\rm m}) \cdot I_r  \cdot(1-E_{pi})$ infections resulting in $F_{pi}\cdot (1{\rm m}) \cdot I_r  \cdot(1-E_{pi})\cdot H_r \cdot (1-HRR_{pi})$ hospitalizations.  Thus, the total number of hospitalizations expected in the unvaccinated test population is 
\begin{equation}\label{defU}
 U = (1-F_{pi}) (1{\rm m})  I_rH_r +  F_{pi} (1{\rm m}) I_r (1-E_{pi}) H_r  (1-HRR_{pi}).
\end{equation}

{\em Deriving $V$.} Next, we assume that every individual in the test population is fully vaccinated.  Since all members are vaccinated, those with prior infection have hybrid protection. Once again, we reason in two stages, first accounting for the protection against infection, then for the protection against hospitalization for infected individuals.   Once again, we assume a fraction $F_{pi}$ of the test population has had a prior COVID-19 infection. 

Then, in the fully vaccinated test population, the expected number of hospitalizations is $(1-F_{pi})\cdot(1{\rm m})\cdot I_r \cdot(1-E_v)\cdot H_r \cdot (1-HRR_v)$ for those without prior infection and $F_{pi}\cdot (1{\rm m}) \cdot I_r \cdot (1-E_h) \cdot H_r\cdot (1-HRR_h)$ for those with hybrid protection. So, the total number of hospitalizations in the population is
\begin{equation}\label{defV} 
  V = (1-F_{pi})(1{\rm m}) I_r(1-E_v)H_r  (1-HRR_v) + F_{pi}  (1{\rm m}) I_r(1-E_h) H_r (1-HRR_h).
\end{equation}

{\em Hospitalizations prevented by vaccination.}  As noted above, the difference $U-V$ is the hospitalizations-prevented benefit of vaccination for the test population:
\begin{multline}\label{e.benefit}
U - V = (1{\rm m})I_rH_r \biggr(1 - (1 - E_v)(1 - HRR_v)(1 - F_{pi}) \\[-10pt]
  - F_{pi}\left(\rule{0in}{0.13in} 1 + (1 - E_h)(1 - HRR_h) -(1-E_{pi})(1 - HRR_{pi})\right)\biggr).
\end{multline}

We remark that disregarding the protection afforded by prior infection, as the FDA did, is equivalent in our model to assuming that all persons in the test population are infection-na\"ive, i.e., that $F_{pi} = 0$. With $F_{pi}=0$ in Equation~(\ref{e.benefit}), its right-hand side is
\begin{equation}\label{Fpi0}
 \left[\text{Expected Baseline Hospitalizations}\right]\left(1-(1-E_v)(1-HRR_v)\right),
\end{equation}
which is essentially the FDA's model $H\cdot VEH$ for hospitalizations prevented. The quantity $1-(1-E_v)\cdot (1-HRR_v)$ of (4) corresponds to the $VEH$ formula on page 10 of \cite{UKTB}, which is the reference the FDA used in \cite{Yogurtcu} for its $VEH$ estimate of 0.72. Thus, our model for hospitalizations-prevented $U-V$ represents a natural extension of the FDA's model.


\subsection{Estimating Values for Model Inputs}\label{ss.inputs}
 Like the FDA, we account for uncertainties in model inputs by adopting a scenario-based  assessment.  The FDA's model-input assumptions for its six scenarios of  \cite{Yogurtcu} are summarized in Table~\ref{MIA} above. Observe that the first three scenarios differ only in their COVID incidence assumptions, with Scenario 1 representing the FDA's most likely scenario.  Our first three scenarios, Scenarios A--C, are reanalyses, respectively, of the FDA's Scenarios 1--3 because our COVID incidence assumptions are derived from the FDA's.   Based on information available at the time of the FDA's mRNA-1273 analysis, we consider the COVID incidence assumption of our Scenario D more likely to be accurate than those of Scenarios A--C; also, for Scenario D, our values for other inputs represent our best estimates.  Thus, Scenario D is our most likely scenario.  In Scenario E, our final scenario, we assume that prior infection and mRNA-1273 vaccination provide equivalent protection against Omicron, which, for reasons discussed Section~\ref{Benefits}, we believe to be the minimum reasonable assumption of prior infection's protectiveness relative to vaccination's protectiveness.  Model inputs and outputs for our scenarios are described in the next section, while the methods we use to estimate input values  are described below and summarized in Table~\ref{InputsTable}. Our estimates are based on data available to the FDA before 22 January 2022.

{\em Estimating $I_r$}.   As we noted in Section~\ref{Benefits}, the FDA's COVID-case-rate assumption for its most likely scenario, Scenario 1, corresponds to the assumption that essentially all unprotected (i.e.,  unvaccinated) 18--25-year-old males become infected over the  5-month evaluation period.    Thus, in Scenario A, our reanalysis of Scenario 1, we too assume all unprotected persons in the test population become infected.  However, in our modeling, the unprotected are the infection-na\"{i}ve unvaccinated.  Assuming all such persons become infected over 5 months is equivalent to setting $I_r= 1$, our Scenario-A value for $I_r$.  As discussed in Section S1 of Supplement S1,  for our Scenarios B and C, values of $I_r$ are determined by the FDA's COVID-incidence assumptions for its Scenarios 2 and 3, respectively, while for our Scenarios D and E, $I_r$ is derived from the assumption that over the evaluation period the increase in COVID-infection level  among 18--25-year-old males is twice that of the general population during the second COVID wave in the U.S. (1 October 2020--28 February 2021). 

  {\em Estimating $H_r$}.   We estimate $H_r$ using an IHR model  \cite{HE} developed via a seroprevalence-based meta-analysis anchored to the ancestral strain of SARS-CoV-2 and available as a preprint in October 2021 \cite{HEP}. Applying the model, we obtain a ``with COVID'' IHR of 0.37\% for the infection-na\"{i}ve unvaccinated of ages 18--25 and reduce this rate assuming 40\% of hospitalizations attributed to COVID are incidental (see Section~\ref{Benefits}), arriving at an IHR of $0.22\% \approx0.6\cdot0.37\%$ (for infection-na\"{i}ve unvaccinated 18--25-year-olds).  The 0.22\% rate for the general population of 18--25-year-olds in the U.S. (males and females) is consistent with U.S. military COVID-hospitalization data \cite{MHD}.  See Supplement S1 for further details concerning our derivation of the IHR value $0.22\%$.

In its benefit--risk assessments \cite{Funk,Yogurtcu} for, respectively, Pfizer's and Moderna's mRNA vaccines, the FDA relied on sex-specific CDC hospitalizations data.  The case and hospitalization rates per 100,000 persons in Table 2 of  \cite{Funk} suggest that for the 18--24 age range, females are 2.53 times as likely as males to be hospitalized with COVID-19 and 1.21 times as likely to have a case of COVID-19.\footnote{These relationships are consistent with those of \cite[Table 1]{Kaim} based on data from 1 March 2020  to 1 March 2022.}   In Section S2 of Supplement S1, we use hospitalization and case data from Table 2 of \cite{Funk} to estimate the IHR for unvaccinated, infection-na\"{i}ve males 18--25 to be $0.14\%$; thus, we set $H_r = 0.0014$.   This value agrees with a natural IHR estimate derived from the FDA's BLA-Memo scenario (discussed in Section~\ref{BLAScenario}).  For the BLA-Memo Scenario, vaccination prevents 76,362 cases and 1755 hospitalizations by providing 30\% effectiveness against cases and 72\% effectiveness against hospitalizations. Thus, the scenario assumes that $76,362/0.3 =  254,540$ cases among the unvaccinated would lead to $1755/0.72 \approx 2438$ hospitalizations, yielding a case hospitalization rate of $\approx$0.958\%.   Applying the CDC case-to-infection multiplier of 4 and assuming a 40\% incidental COVID-hospitalization rate, we arrive at the estimate \mbox{$H_r = 0.6 \cdot (2438)/(4\cdot$ 254,540) $\approx$ 0.0014.}  

{\em Estimating $F_{pi}$}.  For our Scenarios A--C, we use the value $F_{pi}=0.62$, which is the average of (a) the CDC's estimate, mentioned earlier, that 54.9\% of those 18--49 years old had been infected by COVID-19 before 1 October 2021~\cite{CDC1} with (b) an estimate derived in Supplement S1 that 69\% of unvaccinated male 18--25-year-olds had been infected by 1 January 2022.   For Scenarios D and E, we use $F_{pi} = 0.69$.    These values of $F_{pi}$ may be underestimates  because they rely on the CDC's infected rate of 54.9\% by 1 October 2021 for the general population of \mbox{18--49 year-olds}, and we would expect a somewhat higher infected rate among the unvaccinated.  
   
  {\em Estimating $E_v$, $HRR_v$}.   Recall that the FDA used $VE = 0.30$ and $VEH = 0.72$ for all its Omicron-based scenarios \cite[Table 1]{Yogurtcu}.  For reasons discussed in Supplement S1, for all our scenarios, we use $E_v = 0.30$ and $HRR_v = 0.67$. Hence, we are assuming a $VEH$ of  $1- (1-0.30)(1-0.67)\approx 0.77$, {\it which exceeds the $VEH$ used by FDA}.  

{\em Estimating $E_{pi}$ and $E_{h}$.}  Our values $E_{pi} = 0.45$ and $E_h = 0.57$ are derived (see Supplement S1) from a December 2021 study \cite{DH}, based on data from South Africa.\footnote{As discussed in Supplement S1, our effectiveness estimates involving prior infection are consistent with those of meta-analyses  \cite{BWM} and \cite{IHMEFT}. }

{\em Estimating  $HRR_{pi}$ and $HRR_{h}$.}  In Supplement S1, we use data from \cite{Ferguson} (based on Omicron cases in England for the period 1 December 2021--14 December 2021) to estimate $HRR_{pi} = 0.79$ and $HRR_h = 0.88$.\footnote{The estimates $HRR_{pi} = 0.79$ and $HRR_h = 0.88$ combined with our estimates $E_{pi} = 0.45$ and $E_{h} = 0.57$ yield overall estimates for protection against hospitalization---88.5\% for prior infection and 94\% for hybrid protection---consistent with those of meta-analyses  \cite{BWM} and \cite{IHMEFT}.   We note that hybrid protection's exceeding vaccine-only protection  is consistent with studies such as {\cite{Horn, Payne}.}
} .

{\em Estimating $H_{\rm VAM/P}$.}   In Section~\ref{VAMPRisk}, we combined four assessments of VAM/P risk based on data available before 1 January 2022 to obtain an estimate of 250 VAM/P hospitalizations  per million 2nd doses of mRNA-1273 given to 18--25-year-old males.  In Section S6.7 of Supplement S1, we discuss VAM/P-hospitalization risk among 18--25-year-old males for dose 1 of mRNA-1273, justifying a choice of 18 VAM/P hospitalizations per million 1st doses, which is the 1st-dose risk the FDA uses in its BLA-Memo Scenario---compare the first two rows of Table~\ref{SevenScenarios} above.   Thus, for all our scenarios, we set $H_{\rm VAM/P} = 268$.\footnote{In Supplement S1 (Appendix S2, Section S2.4), we show that our VAM/P hospitalization-rate estimate of 268 per million full mRNA-1273 vaccinations among 18--25-year-old males is consistent with VAM/P data of three major studies \cite{Karlstad, Vu, Naveed} appearing in 2022 but available only after 22 January 2022, showing the data suggests a VAM/P-hospitalization rate between 240 and 277 per million full, homologous mRNA-1273 vaccinations among males 18–25.} 

\begin{table}[h!] {\fontsize{10}{11}\selectfont
\begin{adjustwidth}{-1cm}{}
\captionsetup{width=1.2\textwidth}
\caption{\small  Summary of Input Variables*\rule{4.45in}{0in}} \label{InputsTable}
\vspace{-.1in}

\begin{tabular}{|c|c|c|}
\hline
\rowcolor{mygray}
Variables for $U$ and $V$ & Range of Values & Scenario A Values\\ \hline
$I_r$: \parbox[t]{4.4in}{Baseline-infection rate—fraction of an unvaccinated, infection-na\"{i}ve test-population of 18--25-year-old males expected to contract an infection over the 5-month evaluation period. Infections need not accumulate at a constant rate \rule[-3pt]{0in}{.1in}}  & \rule[-40pt]{0in}{.7in}  $0 \le I_r \le 1$ & $1$ \\  \hline
 $H_r$: \parbox[t]{4.5in}{Baseline infection-hospitalization rate for infection-na\"{i}ve unvaccinated males 18--25 \rule[-3pt]{0in}{.1in}}& \rule[-12pt]{0in}{.3in}  $0 \le H_r \le 0.01$ & $0.0014$ \\ \hline
$F_{pi}$: \parbox[t]{4.2in}{ Fraction of a test population of 1 million unvaccinated males 18--25 having had a COVID-19 infection before the start of the 5-month evaluation period \rule[-3pt]{0in}{.1in}}   &  \rule[-12pt]{0in}{.3in} $0 \le F_{pi} \le 1$ & { 0.62} \\ \hline
$E_{pi}$: \parbox[t]{4.2in}{Effectiveness of prior-infection protection against  COVID-19 infection }  & $0 \le E_{pi} \le 1$ & { 0.45} \\ \hline
$E_{v}$: \parbox[t]{4.2in}{ Effectiveness of vaccination against  COVID-19 infection }  & $0 \le E_{v} \le 1$ & $0.3$ \\ \hline
$E_{h}$: \parbox[t]{4.2in}{Effectiveness of hybrid protection against  COVID-19 infection}  & $0 \le E_{h} \le 1$ & { 0.57} \\ \hline
$HRR_{pi}$: \parbox[t]{4.2in}{Hospitalization-risk reduction for those reinfected and unvaccinated} & $ 0 \le HRR_{pi}  \le 1$ & $0.79$  \\ \hline
$HRR_v$: \parbox[t]{4.2in}{Hospitalization-risk reduction for those fully vaccinated and experiencing their first COVID-19 infection }\rule[-15pt]{0in}{.34in}  & $0 \le HRR_v \le 1$ & $0.67$ \\ \hline
$HRR_h$: \parbox[t]{4.2in}{Hospitalization-risk reduction for those reinfected and  fully vaccinated } & $0 \le HRR_h \le 1$ & $0.88$ \\ \hline
{$H_{\rm VAM/P}$}:  \parbox[t]{4.2in}{  Projected number of VAM/P hospitalizations occurring in the course of attaining a test population of 1 million fully mRNA-1273 vaccinated 18--25-year-old males }\rule[-15pt]{0in}{.34in}  & \rule[-25pt]{0in}{.45in} $H_{\rm VAM/P} \approx 268$ & $268$ \\ \hline
\end{tabular} \\
\rule{0in}{0in}\rule{0in}{.1in} \footnotesize{ \hspace{-.1in}\parbox{1.1\textwidth}{\footnotesize * Remarks.  All measures of effectiveness of protection against infection ($E_{pi}$, $E_v$, and $E_h$) are relative to infection-na\"{i}ve, unvaccinated persons.  All hospitalization-risk reductions ($HRR_{pi}$, $HRR_v$, and $HRR_h$) are relative to unvaccinated persons experiencing primary infections.} }
\end{adjustwidth}}
\end{table}

\section{Reanalysis: Results}\label{ResultsSec}

    In Table~\ref{ResultsTable} below, we  compare the hospitalizations-prevented benefit of mRNA-1273 vaccination to corresponding VAM/P-hospitalizations risk, with  Scenarios A--C being reanalyses of the corresponding FDA Scenarios 1--3. \pagebreak

 \begin{table}[h] {\fontsize{10}{11}\selectfont
\begin{adjustwidth}{-1.2cm}{}
\captionsetup{margin={0 cm,  -1 cm}}

\caption{\small Scenarios comparing the  hospitalizations-prevented benefit of mRNA-1273 vaccination of a hypothetical test population of 1 million males 18--25  to the  corresponding VAM/P hospitalization risk $H_{\rm VAM/P}$.  For the FDA's Scenarios 1--3,  the values of $U$ and $V$ are determined by the FDA's model  described in Section~\ref{Benefits}, while, for Scenarios A-E, the values of $U$ and $V$ are determined .by our model described in Section~\ref{MethodsSec}. A green background indicates benefit exceeds risk.   }\label{ResultsTable}
\vspace{-.1in}

\def\arraystretch{0.9}
\begin{tabularx}{1.14\textwidth}{|c|c|c|X|X|c|c|c|}
\hline
\rowcolor{mygray}
 \hspace{-.1in} Scenario  \hspace{-.1in} &  \hspace{-.1in} Description  &  \hspace{-.1in} Assumptions*&   $U$ & $V$& \hspace{-.11in} \begin{tabular}{c}Hospitalizations\\ Prevented \\Per Million\\ Vaccinations\\$U-V$\end{tabular} \hspace{-.15in} &  \hspace{-.1in} $H_{\rm VAM/P}$ \hspace{-.15in}&  \hspace{-.11in} \begin{tabular}{c} Benefit\\Risk \\Ratio \end{tabular} \hspace{-.12in} \\ \hline
\rowcolor{SpringGreen}
\hspace{-.07in}\begin{tabular}{c} FDA\\ Scenario 1\end{tabular}\hspace{-0.1in} & \rule[-23pt]{0in}{0.74in} \hspace{-.11in} \begin{tabular}{l} Highest 2021 COVID \\incidence assumed.\\ FDA's ``most likely\\ scenario''\rule[-10pt]{0in}{0.1in} \end{tabular}  \hspace{-.15in} &  \begin{tabular}{c}  $U = 6619$\ \\ $V$ = 6619(1-0.72) $\approx$ 1853\end{tabular} \vspace{-.09in}
& 6619  & 1853 & 4766 &  \hspace{-.11in} \begin{tabular}{c} 110\\(Dose\,2 only)\end{tabular}  \hspace{-.15in} &43.33 \\  \hline
\rowcolor{SpringGreen}
A \rule{0in}{0.15in} & \rule[-45pt]{0in}{.7in}   \parbox[t]{1.05in}{Reanalysis of\\ FDA's Scenario 1}
&  \parbox[t]{1.65in}{$I_r=1, H_r=0.0014, \\F_{pi}=0.62,  E_{pi}= 0.45,\\  E_v = 0.30,  E_h=0.57,$\\  $HRR_{pi}$=0.79,$HRR_v$=0.67,\\ $HRR_h = 0.88$}& 632 & 168 & 464 & 268 &1.73 \\  \hline
\rowcolor{SpringGreen}
\hspace{-.07in}\begin{tabular}{c} FDA\\ Scenario 2\end{tabular}\hspace{-0.1in} &\rule[-20pt]{0in}{0.5in}  \begin{tabular}{l} Average\,2021\,COVID\\incidence assumed\end{tabular}\vspace{-0.05in}  \hspace{-0.1in} &   \rule[-10pt]{0in}{0in}    \begin{tabular}{c}\rule{0in}{0.12in}$U$ = 2900\  \\ $V$= 2900(1-0.72) $\approx$ 812\end{tabular} \vspace{-.1in}
&  2900  &812 & 2088 &  \hspace{-.11in} \begin{tabular}{c} 110\\(Dose\,2 only)\end{tabular}  \hspace{-.15in} & 18.98 \\  \hline
\rowcolor{mypink2}
B\rule{0in}{0.15in} & \parbox[t]{1.05in}{Reanalysis of\\ FDA's Scenario 2}
&  \parbox[t]{1.65in}{$I_r=0.356$, $H_r=0.0014$, \\$F_{pi}=0.62$,  $E_{pi}= 0.45$,\\  $E_v = 0.30$,  $E_h=0.57$,\\  $HRR_{pi}$=0.79,$HRR_v$=0.67,\\ $HRR_h = 0.88$} &  225  & 60 & 165& 268 & 0.62 \\   \hline
\rowcolor{SpringGreen}
\hspace{-.07in}\begin{tabular}{c} FDA\\ Scenario 3\end{tabular}\hspace{-0.1in} &\rule[-20pt]{0in}{0.5in}  \begin{tabular}{l} Lowest\,2021\,COVID\\incidence\,assumed\end{tabular} \vspace{-0.15in} \hspace{-0.1in} & \rule[-10pt]{0in}{0in}  \begin{tabular}{c}\rule{0in}{0.11in}$U$ = 882 \\ $V$= 882(1-0.72) $\approx$ 247\end{tabular}
&  882  &247 & 635 &  \hspace{-.11in} \begin{tabular}{c} 110\\(Dose\,2 only)\end{tabular}  \hspace{-.15in} & 5.77 \\  \hline
\rowcolor{mypink2}
 C\rule{0in}{0.15in} &  \rule[-45pt]{0in}{.7in}  \parbox[t]{1.05in}{Reanalysis of\\ FDA's Scenario 3}
& \parbox[t]{1.65in}{$I_r = 0.052$, $H_r = 0.0014$,\\ $F_{pi} = 0.62$, $E_{pi}= 0.45$,\\ $E_v = 0.30$,  $E_h=0.57$,\\  $HRR_{pi}$=0.79,$HRR_v$=0.67,\\ $HRR_h = 0.88$} & 33 & 9 & 24 &268 &0.09\\  \hline
\rowcolor{mypink2}
 D\rule{0in}{0.15in} & \rule[-45pt]{0in}{.7in} \parbox[t]{1.21in}{COVID incidence \rule{.1in}{0in} twice that for the 2nd COVID wave in the U.S. Our ``most likely scenario''}
& \parbox[t]{1.65in}{$I_r = 0.456$, $H_r = 0.0014$,\\ $F_{pi} = 0.69$, $E_{pi}= 0.45$,\\ $E_v = 0.30$,  $E_h=0.57$,\\  $HRR_{pi}$=0.79,$HRR_v$=0.67,\\ $HRR_h = 0.88$} & 249 & 68 & 181  & 268& 0.68 \\  \hline
\rowcolor{mypink2}
 E\rule{0in}{0.15in} & \rule[-58pt]{0in}{.86in} \parbox[t]{1.21in}{Equivalent protection for prior infected and COVID-na\"{i}ve vaccinated;  incidence as in Scenario D. }
& \parbox[t]{1.65in}{$I_r = 0.456$, $H_r = 0.0014$,\\ $F_{pi} = 0.69$, $E_{pi}= 0.30$,\\ $E_v = 0.30$,  $E_h=0.57$,\\  $HRR_{pi}$=0.67,$HRR_v$=0.67,\\ $HRR_h = 0.88$} & 300 & 68 & 232  & 268& 0.87 \\  \hline
\end{tabularx} 
 \noindent{\footnotesize{\parbox{1.15\textwidth}{*  {  For discussions of ``Assumptions,''  see Section~\ref{ss.inputs} as well as Supplement S1.}} }}
\end{adjustwidth}}
\end{table}

    We illustrate our model's application by carrying out Scenario-A computations. Using Scenario A assumptions from Table~\ref{ResultsTable} and our definitions (1) and (2) of $U$ and $V$, we obtain\\[4pt]
$
U=\left(1-0.62\right)\cdot\left(1{\rm m}\right)\cdot1\cdot0.0014+0.62\cdot\left(1{\rm m}\right)\cdot1\cdot\left(1-0.45\right)\cdot0.0014\cdot\left(1-0.79\right)\approx632, {\rm and}\\[2pt]
$
\noindent\resizebox{\textwidth}{\height}{$
V=\left(1-0.62\right)\cdot\left(1{\rm m}\right)\cdot1\cdot\left(1-0.30\right)\cdot0.0014\cdot\left(1-0.67\right)+0.62\cdot1\cdot\left(1-0.57\right)\cdot0.0014\cdot\left(1-0.88\right) \approx 168.$}\\[4pt]
Thus, for Scenario A, the hospitalizations-prevented benefit of vaccination is projected to be $U-V =\ 632-168=464$ per million full mRNA-1273 vaccinations,  roughly one-tenth of the FDA's Scenario-1 estimate 4766  \cite{Yogurtcu}.  Although Scenario A has a benefit--risk ratio above 1, its assumption that $I_r=1$ makes this scenario highly unlikely.\footnote{Based on seroprevalence and modeling information, we estimate $I_r\le 0.5$ [Supplement S2, Section S2.3].}  For the more likely scenarios B, D, and E, our modeling suggests that  mRNA-1273 vaccination of 18--25-year-old males generated between 16\% and 62\% more hospitalizations for VAM/P alone compared to COVID hospitalizations prevented. To be conservative, we adjust these estimates downward.

As discussed in Section~\ref{VAMPRisk}, for our main analysis (whose results are presented in Table~\ref{ResultsTable} above),  we have included 1st dose VAM/P risk without accounting for 1st dose benefits (which is the choice the FDA made in its BNT162b2 assessment \cite{Funk} and its BLA-Memo assessment of mRNA-1273 \cite{FDABLACR}).  This may be viewed as a conservative choice, being consistent with the principle that vaccines should have ``highly favorable benefit risk profiles''~\cite{Arlegui}.   Nevertheless, as we indicated in Section~\ref{VAMPRisk}, including 1st dose VAM/P risk while not including 1st dose benefits potentially biases an assessment against vaccination.  Thus, we adjust our results to remove this bias by comparing the hospitalizations-prevented benefit of two doses to the VAM/P risk attributable to dose two only.  Although this adjustment potentially biases our outcomes analysis in favor of vaccination, we, conservatively,  characterize the outcomes of our more likely scenarios in the following way:  comparing our estimated 2nd-dose VAM/P risk at 250 hospitalizations per million to the hospitalizations-prevented benefit of two doses---165 per million in Scenarios B, 181 per million in Scenario D, and 232 per million in Scenario E---we conclude that mRNA-1273 vaccination of 18--25-year-old males generated between $ 8\%$ and $ 52\%$ more hospitalizations for VAM/P  compared to COVID hospitalizations prevented, with an increase of $38\%$ for Scenario D.  In particular, relative to hospitalizations, risks exceed benefits in both Scenario D, our most likely scenario, and in Scenario E, a ``best-case scenario'' for vaccination incorporating the minimum reasonable assumption of prior-infection protection vs.\ vaccine protection.

An ideal benefit--risk assessment of a two-dose vaccination series would account for the benefits and risks of both doses. We present such an assessment in Section S2 of Supplement S2 and provide some commentary on it in Section~\ref{SASECT} below, where we discuss sensitivity analyses exploring how modifying model inputs impacts the benefit--risk picture.

\subsection{Subpopulation Analyses Based on Prior-Infection Status} \label{SPA}  There is a compelling case for conducting a benefit--risk assessment for the subpopulation of unvaccinated 18--25-year-old males having prior-infection protection because prior infection not only decreases the hospitalizations-prevented benefit of vaccination but also increases the risk of severe adverse events ({e.g.,  those described in \cite{Blumenthal,Yousaf, Lieu, Hanson, Reddy, Sacconi, HC}) after \mbox{vaccination \cite{Mathi, Beatty}}, with \cite{Mathi} reporting prior infection is ``associated with an increased risk of severe side effects leading to hospital care (1.56 [risk ratio] (95\%CI: 1.14–2.12)).''     Moreover, prior infection may increase the risk of VAM/P \cite{PatonePreprint}---see Appendix S4 of Supplement S2 for details.\footnote{Survey results in \cite{Bettinger} also suggest that prior infection may increase  VAM/P risk; e.g., 2.53\% of those having a confirmed SARS-CoV-2 infection before dose 1 of mRNA-1273 reported ``chest tightness/pain'' during the 7-day postdose period \cite[Table 4]{Bettinger} while only 0.54\% of those not having a confirmed infection before dose 1 reported these symptoms; corresponding percentages for dose 2 are 1.78\% and 0.98\% \cite[Table 5]{Bettinger}.}  
  
For the subpopulation of 18--25-year-old unvaccinated males having a COVID infection prior to the evaluation period 1 January 2022--31 May 2022, we can use our model to compute  the expected number of COVID hospitalizations in various scenarios simply by setting $F_{pi} = 1$ while leaving all other inputs unchanged.  For example, with $F_{pi} = 1$ in Scenario A, our model predicts $U \approx 161.7$ hospitalizations per million and $V\approx 72.24$, yielding a hospitalizations-prevented benefit of 89.46 per million vaccinated, far below VAM/P-hospitalization risk ($\approx$268 per million vaccinated).  

  As we discuss in Appendix S3  of Supplement S1, the preceding paragraph's analysis may be reinterpreted in terms of evaluation-period hospitalization risk for a  typical unvaccinated 18--25-year-old male COVID-infected prior to the evaluation period.  For such a male, the total hospitalization risk  of the choice to vaccinate is essentially the sum of his probability of VAM/P hospitalization and his probability of having a breakthrough-case COVID hospitalization during the evaluation period.  Our analysis is summarized in \mbox{Tables \ref{Hriskratio1} and \ref{Hriskratio2}} below in terms of hospitalization-risk ratios, with ratios $<$1 indicating the choice to vaccinate decreases risk and $>$1 indicating it increases risk.

   \begin{table}[h!] {\fontsize{10}{11}\selectfont
\begin{adjustwidth}{-1.7cm}{}
\captionsetup{margin={0 cm, -1 cm}}
\caption{\small Hospitalization-risk ratios for a typical male 18--25 with prior-infection protection:  risk of hospitalization after choosing mRNA-1273 vaccination (VAM/P or severe COVID-breakthrough infection*) divided by risk of COVID-hospitalization after choosing not to vaccinate.  }\label{Hriskratio1}

\vspace{-.1in}
\begin{tabular}{|c|c|c|c|}
\hline
\rowcolor{mygray}
\begin{tabular}{c}Probability\\ of Infection $I_r$**\end{tabular} &\begin{tabular}{c} Probability of\\ COVID-19 Hospitalization\\ Without mRNA-1273 Vaccination\end{tabular}  & \begin{tabular}{c} Probability of\\ COVID-19 Hospitalization\\ After mRNA-1273 Vaccination\end{tabular} & \begin{tabular}{c} Hospitalization Risk Ratio\\  vaccinated/unvaccinated, with\\ VAM/P Risk  268 in 1 million \end{tabular} \\ \hline
 $1$ & 161.70 in 1 million & 72.24 in 1 million &\rule[-5.5pt]{0in}{0.22in} $2.10 \approx \frac{268 + 72.24}{161.70}$ \\  \hline 
  $0.5$ & 80.85 in 1 million & 36.12 in 1 million & \rule[-5.5pt]{0in}{0.22in} $3.76 \approx \frac{268 + 36.12}{80.85}$\\ \hline
   $0.25$ &40.43 in 1 million & 18.06 in 1 million  &\rule[-5.5pt]{0in}{0.22in} $7.08\approx \frac{268 + 18.06}{40.43}$ \\ \hline
     \end{tabular} \\
     \noindent{
     \parbox{1.15\textwidth}{\footnotesize{\ *COVID hospitalization risk is considered over the 5-month evaluation period\\ \ **$I_r(1-E_{pi})$ is the probability the ``typical male'' (18--25 with prior-infection protection) will become COVID infected over the evaluation period if unvaccinated; e.g., 0.55 if $I_r = 1$}}} 
\end{adjustwidth}}
\end{table}

According to Table~\ref{Hriskratio1}, our modeling suggests that for a typical unvaccinated \mbox{18--25-year-old} male with prior infection, the choice to vaccinate created a hospitalization risk over twice that associated with the choice not to vaccinate even if $I_r =1$ (evaluation-period infection is certain for an infection-na\"{i}ve male 18--25).   Because a ``typical male'' has an ``average  comorbidities level,'' the information in Table~\ref{Hriskratio1}  is difficult to apply. However, if our typical  unvaccinated male of Table~\ref{Hriskratio1} also has no comorbidities, then his  hospitalization risk ratios {\em should exceed those of Table~\ref{Hriskratio1}},\footnote{Compare, e.g., the BMI 23--24 column of Table~\ref{Hriskratio3} to the final column in Table~\ref{Hriskratio1}.}  because risk-ratio denominators should decrease given no comorbidities and there is no evidence that being free of comorbidities reduces VAM/P risk \cite{GOCCSA}.\footnote{ See also  \cite[p.\ 383]{Morgan}.} 

  For an 18--25-year-old male without prior COVID infection, hospitalization-risk ratios become favorable for vaccination:

 \begin{table}[h!] {\fontsize{10}{11}\selectfont
\begin{adjustwidth}{-1.7cm}{0cm}

\captionsetup{margin={0cm, -1cm}}

\caption{\small Hospitalization-risk ratios for a typical male 18---25 {\it without} prior-infection protection:  risk of hospitalization after choosing mRNA-1273 vaccination (VAM/P or severe COVID-breakthrough infection*) divided by risk of COVID hospitalization after choosing not to vaccinate. } \label{Hriskratio2}

\vspace{-.1in}
\begin{tabular}{|c|c|c|c|}
\hline
\rowcolor{mygray}
\begin{tabular}{c}Probability\\ of Infection $I_r$\end{tabular} &\begin{tabular}{c} Probability of\\ COVID-19 Hospitalization\\ Without mRNA-1273 Vaccination\end{tabular}  & \begin{tabular}{c} Probability of\\ COVID-19 Hospitalization\\ After mRNA-1273 Vaccination\end{tabular} &  \begin{tabular}{c} Hospitalization Risk Ratio\\  vaccinated/unvaccinated, with\\ VAM/P Risk  268 in 1 million \end{tabular}  \\ \hline
 $1$ & 1400 in 1 million & 323.4 in 1 million &\rule[-5.5pt]{0in}{0.22in} $0.42 \approx \frac{268 + 323.4}{1400}$ \\  \hline 
  $0.5$ & 700 in 1 million & 161.7 in 1 million &\rule[-5.5pt]{0in}{0.22in} $0.61 \approx \frac{268 + 161.7}{700}$\\ \hline
   $0.25$ &350 in 1 million & 80.85 in 1 million  &\rule[-5.5pt]{0in}{0.22in} $1.00 \approx \frac{268 + 80.85}{350}$\\ \hline
     \end{tabular}\\ 
    \noindent{ \parbox{1.15\textwidth}{\footnotesize{\ *COVID hospitalization risk is considered over the 5-month evaluation period.}}}
\end{adjustwidth}}
\end{table}

\subsection{Subpopulation Analyses Based on Prior Infection Status and BMI} \label{SPABMI}    At this point,  we discontinue confining to footnotes discussions and citations of data unavailable at the time of the FDA's benefit-risk analysis of mRNA-1273.

The FDA acknowledges, ``The health condition of individuals may remarkably impact the B-R profile and its evaluation'' \cite[Section~4]{Yogurtcu}. Comorbidities such as obesity may triple COVID-hospitalization risk \cite{CDCobesity}.  Our model can produce a benefit--risk analysis based on comorbidity status by appropriately adjusting its infection-hospitalization-rate input $H_r$.   We illustrate the process in  Appendix S3 of Supplement S1, focusing on obesity because evidence suggests that Body Mass Index (BMI)  is the single most potent modifier of COVID-19 risk for those under 60 years old  \cite{Tartof,Nagy}.  We use an assessment of COVID-hospitalization risk based on BMI that finds risk is minimized when BMI $\approx 23$ and that for $23 \le \text{BMI} \le 44$ risk is multiplied by a factor of approximately 1.09 (1.08, 1.10) for each unit increase in BMI~\cite{Gao}.  The results of our modeling appear in  Table~\ref{Hriskratio3} below.\pagebreak

 \begin{table}[h!] {\fontsize{10}{11}\selectfont
\caption{Hospitalization-risk ratios $\left(\frac{\text{risk if vaccinated (VAM/P \& breakthrough)}}{\text{risk if unvaccinated}}\right)$ for males 18--25 based on BMI . Ratios displayed as  pairs:
{\small ``with prior-infection protection, without prior-infection projection.''} } \label{Hriskratio3}

\vspace{-.1in}

\begin{tabular}{|c|c|c|c|c|c|}
\hline
\rowcolor{mygray}
\begin{tabular}{c}Probability\\ of Infection $I_r$*\end{tabular} & BMI 23--24  & BMI 25--29& BMI 30--34 & BMI 35--39 & BMI 40--44\\ \hline
 $1$ & 2.56, 0.39 & 1.79, 0.30 & 1.25 , 0.23 & 0.89, 0.19 &0.66, 0.17  \\  \hline 
  $0.5$ & 4.90, 0.66 & 3.36, 0.48 & 2.26, 0.35 & 1.55,  0.27   & 1.09, 0.22  \\ \hline
   $0.25$ &9.56 ,1.20 & 6.49 , 0.84 &4.30 , 0.59 & 2.85, 0.42 &1.85, 0.32\\ \hline
     \end{tabular} \\
     \noindent{\footnotesize{\rule{0.4in}{0in}  *With prior-infection protection, $I_r(1-E_{pi})$ is the probability of infection over the evaluation period.  }} }
\end{table}

  Tables \ref{Hriskratio1} and \ref{Hriskratio2} of the preceding section establish the expectation that for a ``typical male'' 18--25 (having an ``average  comorbidities level''), prior-infection status determines whether vaccination is beneficial.  Consistent with this expectation, Table~\ref{Hriskratio3} above indicates that for those having BMI in the intermediate range 25--34, prior-infection status does determine whether vaccination is beneficial.  However, for those with  high BMI (35--44) in a high-infection-risk scenario, Table~\ref{Hriskratio3} indicates that vaccination becomes beneficial even for those with prior-infection protection.

\subsection{Sensitivity Analyses}\label{SASECT}

Perhaps the most important of our sensitivity analyses assesses how the inclusion of 1st-dose benefits in our Scenarios B--E changes the corresponding benefit--risk ratios. This analysis, carried out in Section S2 of Supplement S2,  yields the following benefit--risk ratios for our more likely scenarios: 0.68 for B, 0.73 for D, and 0.96 for E. Estimating inputs for 1st-dose benefit modeling was especially challenging given the extremely limited data on 1st-dose benefits for Omicron available before 22 January 2022.  Our having lower confidence in { input estimates} for 1st-dose modeling contributed to our decision to present the modeling only as a sensitivity analysis.  
 
 Also, in Section S2 of Supplement S2, we provide a systematic sensitivity analysis related to Scenario D, our most likely scenario.  For each input, we hold all others constant at their Scenario D values, determining the extent of possible variation in the chosen input before equipoise is reached.    For example, if we fix our model input values, except for $I_r$, at their Scenario D levels, then benefit--risk equipoise is reached when $I_r\approx 0.678$, a value nearly 50\% greater than the Scenario D value of $0.456$.  We also point out that the risk-exceeds-benefit conclusion of Scenario D is not sensitive to the value of $E_{pi}$: even if $E_{pi} =0$, the benefit--risk ratio  of this scenario is $\approx 0.83$. The insensitivity of the risk-exceeds-benefit outcome of Scenario D to significant changes in inputs from their Scenario D values increases confidence in the outcome.

\subsection{Model Validation}\label{VTM}    In Supplement S2, we test our model using Ontario COVID-hospitalization data from the evaluation period of 1 January 2022--31 May 2022, showing its projections of the hospitalizations-prevented benefit of vaccination are more accurate than those of the FDA \cite{Yogurtcu}. We  use Ontario data for many reasons; e.g., unlike the CDC, Public Health Ontario chose to publish incidental COVID-hospitalization rates (starting 8 January 2022)  \cite{PHO_IH}; see [Supplement S2, Section S3] for other reasons.  

 We test our model in two ways: (i) showing that it fits Ontario COVID-hospitalization data with inputs at plausible levels for Ontario over the evaluation period [Supplement S2, Section S8], and (ii) using Ontario COVID-hospitalization data to predict corresponding data for the U.S. [Supplement S2, Section S7].  
  
    Based on COVID-hospitalization counts from Public Health Ontario as well as corresponding incidental-hospitalization rates,  we estimate that vaccinating one million \mbox{18--25-year-old} Ontarians prevented 152 COVID hospitalizations over the evaluation period---see Table S2.5  of Supplement S2. 
Making various adjustments, e.g., accounting  for the better health of Ontarians vs.\ Americans, we obtain an {\it upper-bound} estimate that 214 evaluation-period hospitalizations were prevented by one million full COVID vaccinations of 18--25--year-old males in the U.S.---see Table S2.7 of Supplement S2 and the discussion that follows it.  Observe that 214 is an upper bound for our projections of Scenarios B--D (see Table~\ref{ResultsTable}) and a little below our projection 232 of Scenario E,  but all of the FDA's projections, ranging from 635 to 4766 for Omicron-based scenarios, substantially exceed 214.

     Ontario data also supports the definitiveness of the results of Table~\ref{Hriskratio1} above given the male  has no comorbidities, suggesting that 151 in 1 million is an overestimate of his evaluation-period COVID hospitalization risk [Supplement S2, Section S2.3].

\section{Discussion}  \label{Discussion}

By accounting for prior-infection protection and using evidence-based model inputs, we find in our Scenarios B--E of Table~\ref{ResultsTable} that risks of mRNA-1273 vaccination outweighed benefits for the general population of 18--25-year-old males, assuming hospitalizations are the principal endpoint of analysis. In particular, for our scenarios predicting plausible infection-level increases  over the evaluation period (Scenarios B, D, and E), our model suggests that vaccination of 18--25-year-old males generated between 16\% and 62\% more hospitalizations for vaccine-attributable myo/pericarditis (VAM/P) alone compared to COVID hospitalizations prevented, or, discounting 1st-dose VAM/P risk,  between 8\% and 52\% more hospitalizations. Recalling that the FDA found, ``[T]he benefits of vaccination with the Moderna COVID-19 vaccine clearly outweigh the risks, even among the male adolescent population'' \cite[Section~4]{Yogurtcu}, we have demonstrated that the conclusions of a benefit--risk assessment for vaccination may be dramatically impacted by accounting for the benefits conferred by prior infection.  We have also demonstrated in Tables~\ref{Hriskratio1}--\ref{Hriskratio3}  how the benefit--risk picture for vaccination may vary based on an individual's prior-infection and comorbidity status.   Finally, we have demonstrated how a benefit--risk assessment for a two-dose vaccination series may acknowledge potential biases relating to the inclusion or exclusion of 1st dose benefits or~risks.  
 
Consistent with our model's finding that for 18--25-year-old males in the U.S., risks of mRNA-1273 vaccination outweighed benefits relative to hospitalizations, we note that  France suspended mRNA-1273 use on 15 October 2021, making it available again on 8 November 2021 but at half-dose and {\it only for those over 30 years} \cite[p.\ 4]{Vu2}. Also observe our model's finding that vaccination of 18--25-year-old males generated between 8\% and 52\% more hospitalizations for vaccine-attributable myo/pericarditis (VAM/P) compared to COVID hospitalizations prevented is consistent with our model-validation work discussed in Section~\ref{VTM}:  recall that we estimate that at most 214 evaluation-period hospitalizations were prevented by one million mRNA-1273 vaccinations of 18--25-year-old males in the U.S. [Supplement 2, Table S2.7], and note that our corresponding VAM/P-hospitalization projection of 268 is a little over 25\% higher than 214}. 

We emphasize that the cardiac risk of COVID infection is incorporated into our modeling because our estimates of COVID hospitalizations include those for  COVID-related cardiac issues.  We address here assertions such as ``Importantly, the risk of myocarditis after any dose of COVID-19 vaccine, in any population, is substantially lower than the risk of myocarditis after SARS-CoV-2 infection,'' which appeared, without supporting references, in an April 2024 editorial  of the {\it European Heart Journal} \cite{Cooper}. Contrary to the editorial’s assertion, Patone et al.\ find that, ``In men younger than 40 years old, the number of excess myocarditis events [leading to hospitalization] per million people was higher after a 2nd dose of mRNA-1273 than after a positive SARS-CoV-2 test (97 [95\% CI, 91–99] versus 16 [95\% CI, 12–18])'' \cite{Patone}. Karlstad et al.\ also find significantly higher risk of hospitalized myocarditis after dose 2: 183.9 per million 2nd mRNA-1273 doses \cite[Table 2]{Karlstad}  vs.\ 13.7 per million positive tests \cite[eTable 7]{Karlstad}  among males 16–24. Moreover, these myocarditis-incidence rates expressed in cases per million positive tests overstate the risk of myocarditis after COVID infection expressed in cases per million COVID infections (because many infections are not confirmed by positive tests, but essentially all infections among  those seeking care for myocarditis are confirmed by positive tests) \cite{BP}.  Consistent with \cite{BP}, the authors \mbox{of \cite{StoweMyo}} note, “[T]he attributable risk estimates for COVID-19 [associated myocarditis] used laboratory confirmed cases as the denominator and will be affected by the proportion of all SARS-CoV-2 infections captured by testing, precluding a direct comparison with vaccine-associated attributable risks''.

    \subsection{The Strength of Prior-Infection Protection} 
As noted in Section~\ref{Benefits}, at the time of the FDA's mRNA-1273 assessment,  immunological expectations and studies such as \cite{Shenai, CDCSB, Leon, Cohen, DH, Ferguson} suggested that prior-infection protection would likely equal or exceed vaccine protection among the COVID-na\"{i}ve.   In our modeling,  we have assumed that vaccination's effectiveness against hospitalization is 77\% in all scenarios, while that of prior infection is 88.5\% except in Scenario E.
 
   A systematic review \cite{Feikin} involving 21 vaccine-effectiveness studies from the period 3 December 2021--7 April 2022 reports,
   
\begin{quotation}{\noindent In contrast to vaccine effectiveness against delta severe disease, the majority of vaccine effectiveness estimates for omicron severe disease were below 75\%; for example, thirteen (81\%) of sixteen vaccine effectiveness estimates within three months of vaccination with the primary series were below 75\% (Figure). Moreover, 13 (42\%) vaccine-specific estimates fell below 50\% at some point in time after vaccination.}
\end{quotation}

In contrast,  the meta-analysis \cite{BWM} estimates prior infection's effectiveness against severe Omicron to be over 80\% through 6 months after infection, while the \mbox{meta-analysis \cite[Table S2]{IHMEFT}}  estimates its effectiveness to be over 88\% through 45 weeks after infection, with the former estimate based on studies from 1 January 2022--1 June 2022 and the latter, on studies published by 31 September 2022.  In mid-January of 2022,  the minimum reasonable evidence-based assumption would have been that prior-infection protection would equal vaccine protection (as in our Scenario E).

\subsection{Assessing VAM/P Risk for mRNA-1273}  

Contributing to our finding mRNA-1273 vaccination's risks exceeded benefits for 18--25-year-old males relative to hospitalizations is our assessment that VAM/P risk for mRNA-1273 is substantially greater than that suggested by the FDA's BEST System. As detailed in Sections~\ref{VAMPRisk} and \ref{ss.inputs}, our mRNA-1273 VAM/P-risk estimate of 268 hospitalizations per million full vaccinations is based on data, available to the FDA at the time of its mRNA-1273 assessment, from Ontario \cite{BuchanPreprint}, England \cite{PatonePreprint}, and the United \mbox{States \cite{SharffPreprint, Funk, Wong, FDABLACR}}.  In Appendix S2 of Supplement S1, we pool data, unavailable at the time of the FDA's mRNA-1273 assessment, from Scandinavia \cite{Karlstad}, France \cite{Vu}, and British Columbia \cite{Naveed} to obtain a VAM/P-hospitalization rate between 240 and 277 per million full, homologous mRNA-1273 vaccinations among males 18–25.\  A possible explanation for the FDA's underestimating VAM/P risk for mRNA-1273 is that its BEST System has ``inherent limitations, such as small sample sizes and imperfect sensitivity of ICD-10 codes to identify these rare \mbox{outcomes'' \cite[Section~4]{Funk}}.     

Recall that there is evidence that the CDC's VSD system also underascertained VAM/P risk for mRNA-1273.   As noted in Section~\ref{BLAScenario}, during the period  1 June 2021--4 September 2021 of ``enhanced passive surveillance'' among males and females 18–39,  Public Health Ontario's myo/pericarditis detection rate per million 2nd doses of mRNA-1273 was 3.76 times that of the CDC's VSD system based on its data through 9 October 2021 [Supplement S2, Appendix~S3].  The study \cite{Sharff} offers at least a partial explanation for the VSD's lower detection rate by revealing shortcomings in the VSD's search algorithm.

\subsection{Limitations}   Our principal aim has been to illustrate how prior-infection protection, as well as comorbidity status, can be integrated into a deterministic benefit--risk model that extends, in a natural way,  the one the FDA employed in its assessment of mRNA-1273 vaccination.  We have developed and presented our model in accordance with BRIVAC (Benefit--RIsk models applied to VACcines) guidelines, in particular, its item 17 ``Sensitivity/scenario analyses'' \cite{Arlegui2}.  We acknowledge that it is desirable to employ more sophisticated approaches to benefit--risk modeling and to accounting for  uncertainty in such modeling; for instance, Monte Carlo simulation modeling with Beta/Poisson distributions for model inputs, guiding input choices for simulations.

  In applying our model to extend and improve the FDA's mRNA-1273 assessment~\cite{Yogurtcu} relative to its hospitalizations endpoint, we have assumed that young  males are hospitalized for COVID at a lower rate than young females, as did the FDA in \cite{Funk,Yogurtcu}.  Data \mbox{from \cite{Funk,Kaim}} suggests this rate difference is substantial; however, the rates of \cite{Funk,Kaim} include incidental hospitalizations, and incidental rates for males may differ from those for females.  We have assumed, as did the FDA, that vaccination and prior infection offer essentially equivalent protection for males and females.  Also, we have based our assessment of the difference in hospitalization rates between young males and females on data in Table 2 of \cite{Funk} representing mostly pre-Delta cases.  Relative to Omicron, sex-based differences in hospitalization rates might have changed.  However, the rate differences provided by Table 1 of \cite{Kaim}, based on two years of data including that of the initial Omicron wave, are consistent with those of Table 2 of \cite{Funk}.  

COVID-hospitalization data from Ontario [Supplement S2] and from Connecticut [Supplement S1, Appendix S1] suggests that our Omicron IHR value $0.0014$ for 18--25-year-old males is a reasonable estimate for the Omicron IHR for {\em all} 18--25-year-olds (general population mix of males and females). Thus, even if we assume males and females in this age range are hospitalized for COVID at roughly equivalent rates, our IHR value of 0.14\% for 18--25-year-old males appears to be a good estimate.

 In Section S2 of Supplement S1, based on data available before 22 January 2022, we obtain an Omicron IHR estimate for 18--25 year-olds (general-population mix of males and females) of 0.22\%, not 0.14\%.       One possible explanation is that  we may have underestimated the incidental COVID-hospitalization rate for 18--25 year-olds.   Our IHR estimate for \mbox{18--25 year-olds} of 0.22\% results from assuming an incidental Omicron-hospitalization rate of 40\%, a rate suggested by \cite{Walensky,PHO_IH} for hospitalizations with COVID among patients of all ages.   However, as noted in Section~\ref{Benefits} (and discussed in detail in Supplement S1), there is strong evidence that incidental COVID-hospitalization rates are higher among the young than among the elderly.  In particular, COVID-hospitalization data discussed in Appendix S1 of Supplement S1 suggests an incidental rate of over 70\% among 15--24-year-olds in Connecticut during calendar year 2020.   Assuming an incidental rate of about 62.2\% (instead of 40\%) transforms our IHR estimate of 0.22\% into the estimate 0.14\%.    

For a vaccine having high effectiveness against infection that does not quickly wane, vaccination can have an important benefit not captured by our model---providing an enduring reduction in the level of infection in the population that may allow some members to avoid infection entirely through the contribution of vaccination to herd immunity.

Our risk assessment, like the FDA's, ignores known, rare vaccine-related adverse events other than myo/pericarditis that may require hospitalization.  Finally, our risk assessment, like the FDA's, ignores the possibility that COVID-19 vaccination may have long-term health effects; for instance, the FDA's package-insert  information  for Moderna's mRNA-1273 begins, ``SPIKEVAX has not been evaluated for carcinogenic, mutagenic potential, or impairment of male fertility in animals'' \cite[Section~13]{FDAPI}.

\section{Conclusions}

Our model illustrates the importance of nuanced vaccine recommendations based on careful benefit--risk modeling that takes into account health status and prior infection(s), two covariates that were not included in the FDA's models \cite{Funk, Yogurtcu}. For healthy young males with prior infection, our model suggests that risks outweighed benefits (relative to hospitalizations) based on data available prior to licensure of the mRNA-1273 product. However, for higher-risk young males with obesity, and especially those with no prior infection, vaccination was found to confer benefit.

Risks of infection and vaccination, such as long COVID and VAM/P, must be carefully weighed to offer informed consent. Current long-COVID prevalence estimates among adults with confirmed SARS-CoV-2 infection are most reliably estimated in the range of 7--15\% when using standardized definitions and representative sampling \cite{AScot, ShiUS}, and when using a conservative definition consistent with other post-viral syndromes, the WHO finds 2.8\% among those younger than 20 \cite{VosWHO}. 

With respect to VAM/P, more recent formulations of Moderna’s COVID vaccine have a reduced mRNA dose and VAM/P-incidence rates have declined \cite{Hviid}.  Long-term VAM/P prognosis remains somewhat unclear. A CDC follow-up at 90 days found that although the majority (81\%) of VAM/P cases recovered, 54\% had ongoing abnormal cardiac findings and 26\% were on daily cardiac medication; only 68\% were cleared for all physical activity \cite{Kraca}. Clinical guidelines for return-to-sport following myocarditis call for 3--6 months' activity restriction \cite{Yama, Marsch}. In addition, there is the unknown prognosis for complications like late gadolinium enhancement (LGE) \cite{Hadley}.

The question of boosting the general population, including healthy young males, is beyond the scope of this paper and requires tremendous nuance, a fact widely recognized by fall 2021 \cite{Krause}. The FDA indicated in May 2025 \cite{VPra} that it anticipates needing data from RCTs to evaluate clinical outcomes prior to authorizing COVID-19 vaccines for healthy individuals ages 6 months to 64 years, while immunogenicity standards will likely suffice for those with at least one chronic condition and those over age 65. The agency’s move thus operationalizes the risk-stratified approach supported by our deterministic modeling and will generate data that can be used in more sophisticated modeling.

\bigskip

{\small
\noindent{\bf Acknowledgment}: We thank Retsef Levi for providing comments that improved both the structure and content of this paper.

\noindent{\bf Supplements}:
\noindent \href{https://drive.google.com/file/d/16tTPqf5mbLrgHkGVa_WymqAYSn86KbFr/view?usp=share_link}{\color{blue}\underline{Supplement S1}}, 
\href{https://drive.google.com/file/d/1axYvxHeh_8ZxjIVsy2ASvHjhVWjWpC8s/view?usp=share_link}{\color{blue}\underline{Supplement S2}}}

\end{document}